\documentclass[12pt]{article}%\documentclass[preprint,showpacs,preprintnumbers,amsmath,amssymb]{revtex4}

\usepackage{showlabels}
\usepackage{amsfonts}
\usepackage{color}
\usepackage{relsize}
\usepackage{graphicx}% Include figure files
\usepackage{dcolumn}% Align table columns on decimal point
\usepackage{bm}% bold math
\usepackage{wrapfig}
\usepackage{graphicx}

\newcommand{\be}{\begin{equation}}
\newcommand{\ee}{\end{equation}}
\newcommand{\ba}{\begin{array}}
\newcommand{\ea}{\end{array}}

\title{Multiscale   mechanics of macromolecular materials \\ with unfolding domains}

\begin{document}

\date{}

\maketitle \vspace{- 1.5 cm}

\begin{center}
\large{
D.~De Tommasi$^1$,
 G.~Puglisi$^1$, G.~Saccomandi$^2$.}
\end{center}

\small{
\date{} \vspace{ 1cm}
\noindent $^1${Dip. Scienze dell'Ingegneria Civile e dell'Architettura,
 Politecnico di Bari}\vspace{0.2 cm}

\noindent $^2${Dipartimento di Ingegneria Industriale, Universit\`a degli Studi di Perugia}}

\vspace{ 1 cm} 
\begin{abstract}
\noindent 
We propose a general multiscale approach for the mechanical behavior of three-dimensional networks of macromolecules undergoing strain-induced unfolding. Starting from a (statistically based) energetic analysis of the macromolecule unfolding strategy, we obtain a three-dimensional continuum model with variable natural configuration and an energy function analytically deduced from the microscale material parameters. The comparison with the experiments shows the ability of the model to  describe  the complex behavior, with residual stretches and unfolding effects, observed in different biological materials.

\vspace{ 1 cm}


\end{abstract}

\maketitle

\section{Introduction}
 The  phenomenological theory of Continuum Mechanics represents  a recognized and  successful  framework to describe a wide class of material behaviors. In the last decades the scientists working in this field have been involved, for  crucial theoretical and applicative interests, in a significant extension of the methods of non-linear Continuum Mechanics finalized to introduce multiscale analyses `intrinsically' required for the description of the rich and complex behavior of many  artificial and biological materials. Indeed, a deep comprehension of the response of these materials, exhibiting strong nonlinearities, history dependence, growth, healing and softening effects, can be  reached only by analyzing how the macroscopic response results from the complex material behavior at the meso, micro and nano-scales (\cite{BA}, \cite{CMD}, \cite{DeTommasi-Puglisi-Saccomandi}, \cite{DeTommasi-Puglisi-Saccomandi3}, \cite{SpecI}).  
 
 The necessity of such an effective multiscale approach has been particularly evidenced in the field  of biomechanics to explain the complex mechanical behavior of  soft tissues and in the related field of the design of new bioinspired materials \cite{Lv}. Such an effort has brought an extraordinary and successful impulse in the development of different new effective \emph{homogenization} techniques,  deducing macroscopic constitutive laws starting from the  meso/microscopic material properties \cite{SpecI}.  
 
The present paper is inscribed in this research field.  
A wide class of soft (biological and polymeric) materials are constituted by networks of macromolecules with 
 folded ({\it e.g.} $\alpha$-helices or $\beta$-sheets) and unfolded domains.
 Atomic Force Microscopy experiments  
 (see {\it e.g.} \cite{Rief1} for titin,  \cite{BL} for fibrinogen, \cite{SK} for silks, \cite{Smith} for DNA/RNA strands and  \cite{R} for polysaccharides) show that under increasing end-to-end length the macromolecular chains undergo successive events of domains unfolding. These phenomena are revealed by a (typically periodic) sequence of stress drops arising from domains unravelling (see Fig.\ref{modelsoft}). The main physical results of these  unravelling phenomena are a sudden entropy variation, due to the increased number of monomers, and a change of the natural configuration (variable contour length). Macroscopically this phenomenon induces a complex, history-dependent dissipative  behavior with softening, residual stretches and  volume variation. To analytically relate the behavior at the different scales, we consider two different scale changes: a discrete-continuum limit for the single macromolecule behavior and a micro-macro statistically based network  approach to deduce a three-dimensional material law.  \vspace{0.2 cm}

More in detail, 
to describe the behavior of the single macromolecule, we refer to the statistically inspired  approach proposed in \cite{DMPS} that deduces the total (entropic plus unfolding/enthalpic) energy function by considering a force-length relation with a variable contour length. This analysis is summarized in Sect. \ref{ea} where we also deduce an analytic expression of the unfolding force and length based on a (Griffith--type) energy minimization procedure.

In Sect.\ref{cl}, aimed to fully analytical results, we propose a continuum approximation of the macromolecular energy  by considering the case of chains with a large number of domains. Interestingly, in this limit we obtain that the unfolding corresponds to a constant force (force plateaux) and a constant relative stretch $\lambda_{\mbox{\tiny {\it rel}}}^{\mbox{\tiny {\it un}}}$ (ratio between the chain length and  contour length). As we prove in the following sections, $\lambda_{\mbox{\tiny {\it rel}}}^{\mbox{\tiny {\it un}}}$ represents the main parameter regulating the macroscopic three-dimensional response. Then we show that the continuum chain model can be  inscribed in the classical field of thermodynamics with internal variables \cite{Z}, with our internal variable representing the fraction of folded domains. Then we deduce the flow rules and the corresponding dissipation potential and, by analyzing the classical Clausius-Duhem inequality, we prove the thermodynamic consistence of the obtained model.  As we show, the equilibrium behavior of the continuum model can be described by two equivalent variational problems: one minimizing the total (entropic plus unfolding) energy of the chain, the other  maximizing the dissipation. Finally, based on a classical Statistical Mechanics approach \cite{Rubinstein}, we describe the variable natural length (zero force) of the unfolding chain and deduce the resulting decomposition of the chain stretch in elastic and permanent stretch as a function of the internal variable.

In Sect.\ref{3D} we introduce the micro-macro scale passage that let us deduce a  three-dimensional  {\it macroscopic} constitutive law. In particular, aimed again to a fully analytical formulation, we consider the efficient 8-chain Arruda-Boyce scheme \cite{AB}. In this regard, 
we observe that
our approach is general and can be extended to other well-known  micro-macro schemes: other network cells schemes, full-network models, non-affine approaches  \cite{Beatty}, \cite{MGL}, \cite{K}, etc.. 
As a result, we deduce a three-dimensional model  
that  can be inscribed in the field of Growth Mechanics (see \cite{Tab}, \cite{All} and \cite{BG} and reference therein). Specifically,  by assuming the classical multiplicative decomposition for the deformation tensor proposed in \cite{Rodr}, we deduce the macroscopic energy density function defined on a variable natural configuration. This energy function is of a Gent-type \cite{PugSac} with a variable limit threshold depending on the value of the internal variable. 
 
The obtained three-dimensional continuum body is characterized by a volume growth (see \cite{Rodr} and \cite{EM}), an {\it isotropic residual stretch} effect and   an {\it isotropic softening}, that  we can here  predict based on the microstructure material properties.
Also in the three-dimensional case we deduce the flow rules and the dissipation potential of the model
with an explicit dependence of the growth tensor on the maximum past value of the first strain invariant. We then discuss the thermodynamical consistence by analyzing the Clausius-Duhem inequality also in the three-dimensional setting. Interestingly the reduced dissipation inequality involves the Eshelby tensor measuring the energy dissipation due to the volume increase.

In Sect.\ref{DG} we apply our model to the analysis of simple extension, biaxial extension and shear deformations and we deduce the corresponding stress-stretch relations.  In particular  we analyze the influence of the microscopic (macromolecular) parameters on the macroscopic response. The obtained behavior is shown to reproduce many experimental phenomena observed at the macroscale of biological materials. Finally we test our model by quantitatively reproducing different experimental behaviors of biological materials (mouse skin reported in
\cite{Dobl}, spider silks in  \cite{VGS}, and heart tissues in \cite{DSY}).

\section{Micromechanics of single chain unfolding}\label{ea}

To describe the behavior of the single macromolecule we refer to the statistically based energetic approach proposed in  \cite{DMPS} (we refer the reader to this paper for details). We model each chain as a lattice of two states domains that can undergo a {\it hard} (folded) $\rightarrow$ {\it soft} (unfolded) transition under growing length. By introducing a (Griffith-like) energy minimization scheme, 
we search for the global minima of the total (entropic plus unfolding) potential energy of the chain.  Specifically, in order to obtain fully analytical results and focus on the main physical effects of the unravelling phenomenon (see  \cite{DMPS} for more general hypothesis),
we consider the case of identical unfolding domains with the same unfolding energy $Q$ (energy `dissipated' in the folded-unfolded transition) and a constant number of monomers released  in each transition, leading to a constant increase $l_c$ of the contour length $L_c$ of the chain (see {\it e.g.} \cite{Rubinstein}).  Moreover, we neglect (see \cite{DMPS}) the elasticity and the dimension of the folded domains and the mixing (folded/unfolded) entropy.

 Based on these hypotheses,  the total energy $\Phi$ of the chain depends on the number $n$ of unfolded domains and on the  {\it relative chain stretch}
$$\lambda_{\mbox{\tiny {\it rel}}}=\frac{L}{L_c}\in (0,1),$$ 
measuring the elongation as compared with its limit chain extensibility ({\it contour length}) that under the considered hypotheses is given by \begin{equation}L_c=\hat L_c(n)=n l_c , \hspace{1 cm} n\in (n_o,n_t).\label{lcc}\end{equation}
Observe that if $n_o$ denotes the initial  number of unfolded domains and $n_t$  the total number of hard domains, then  $L_c\in(L_c^o,L_c^{\mbox{\tiny 1}})$ where 
\begin{equation}L_c^{\mbox{\tiny 0}}=n_o l_c \label{Lco}\end{equation} is the 
 the initial (unravelled) contour contour length and \begin{equation}L_c^{\mbox{\tiny 1}}=n_t l_c \label{Lu} \end{equation} is  the fully unfolded  contour length.

Speciffically, under previous assumptions we have  (see again \cite{DMPS} for a statistical justification based on an Ising type transition energy)

\begin{equation}
  \Phi=\Phi_e+n Q, \label{llll}
\end{equation}
where $\Phi_e$
is the entropic energy of the unfolded fraction, $n Q$ is the energy expended to unfold $n$  domains. Here we assume
\begin{equation}\label{Phie}
  \Phi_e=
\kappa  \frac{\lambda_{\mbox{\tiny {\it rel}}} ^2}{1-\lambda_{\mbox{\tiny {\it rel}}} }  L_c\end{equation}
and we refer the reader to  \cite{DMPS}  for a detailed discussion of the advantage of this expression of the energy density  (keeping the same asymptotic behavior of the WLC  in \cite{Marko-Siggia}) and for a physical justification of this assumption.  Here
$$\kappa=\frac{k_B\, T}{4\, P},$$
with
 $k_B$  the Boltzmann constant, $T$  the temperature and $P$ the {\it persistence length} of the chain (\cite{Rubinstein}). 
 
By differentiating $\Phi_e$ with respect to $L$, we get the force-length relation\begin{equation}
f= \kappa\frac{2\lambda_{\mbox{\tiny {\it rel}}}-\lambda_{\mbox{\tiny {\it rel}}}^2}{(1-\lambda_{\mbox{\tiny {\it rel}}})^2}. 
\label{WLC}
\end{equation}

 We remark that a variational approach has been fruitfully applied in other rate-independent   dissipative effects such as polymer damage mechanics  (see \cite{DP}, \cite{DMPS2} and references therein) and decohesion problems (see \cite{PT} and references therein). According to this approach, the unfolding events are regulated by the energetic competition between the entropic energy decrease and the enthalpic energy increase  (here assumed as irreversibly lost) associated to the single unfolding event. A more refined analysis would require the study of the wiggly energy landscape and a statistical analysis of the escape times from the multiple energy wells corresponding to the different unfolding configurations. The resulting unfolding strategy depends then also from the rate of deformation \cite{Qi}. Here, aimed again to a fully analytical approach, we consider the classical Maxwell hypothesis (see {\it e.g.} \cite{Bert}) assuming that the configurations of the system corresponds always to the {\it global} minimizers of the total energy $\Phi$. We refer the reader again to \cite{DMPS} (and references therein) for a detailed analysis of the possibility of applying this approach to protein macromolecules unfolding.

\begin{figure}[!h]
\begin{center}$$\begin{array}{rrr}\includegraphics[scale=0.18]{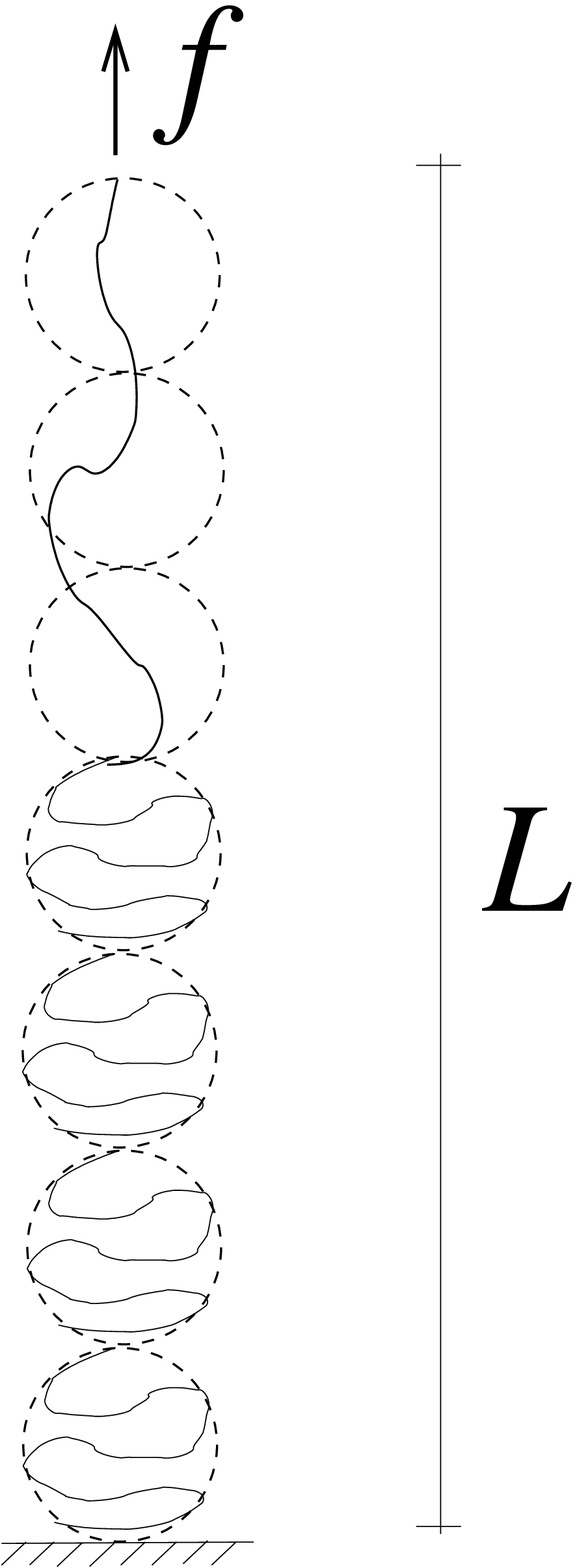} &
\includegraphics[scale=0.57]{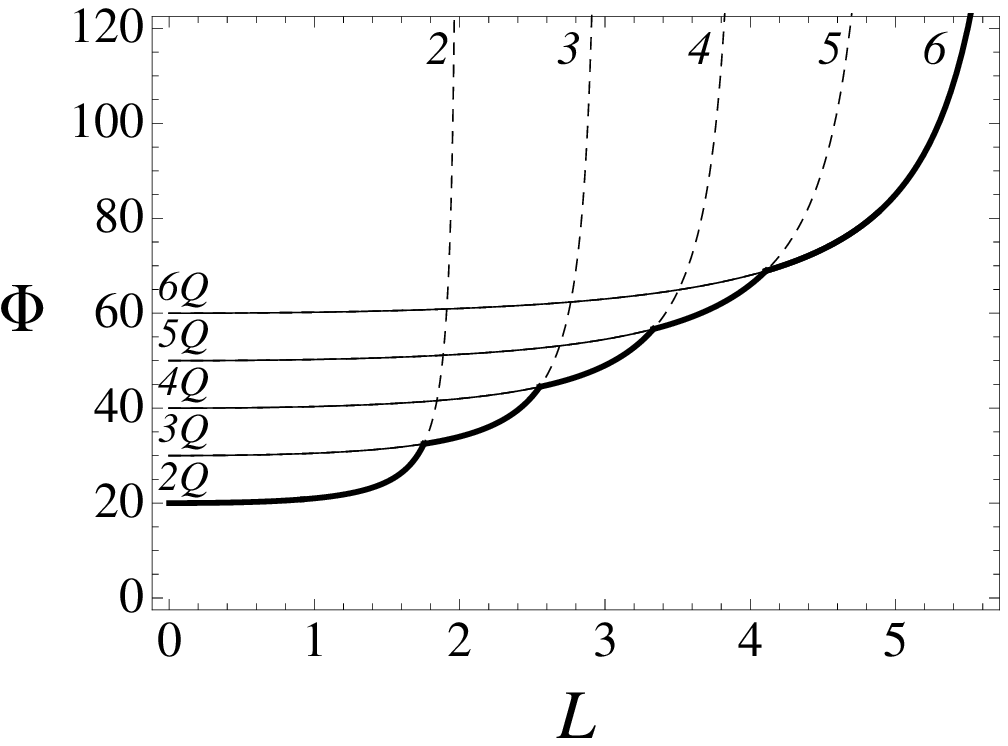} &\, \, \includegraphics[scale=0.57]{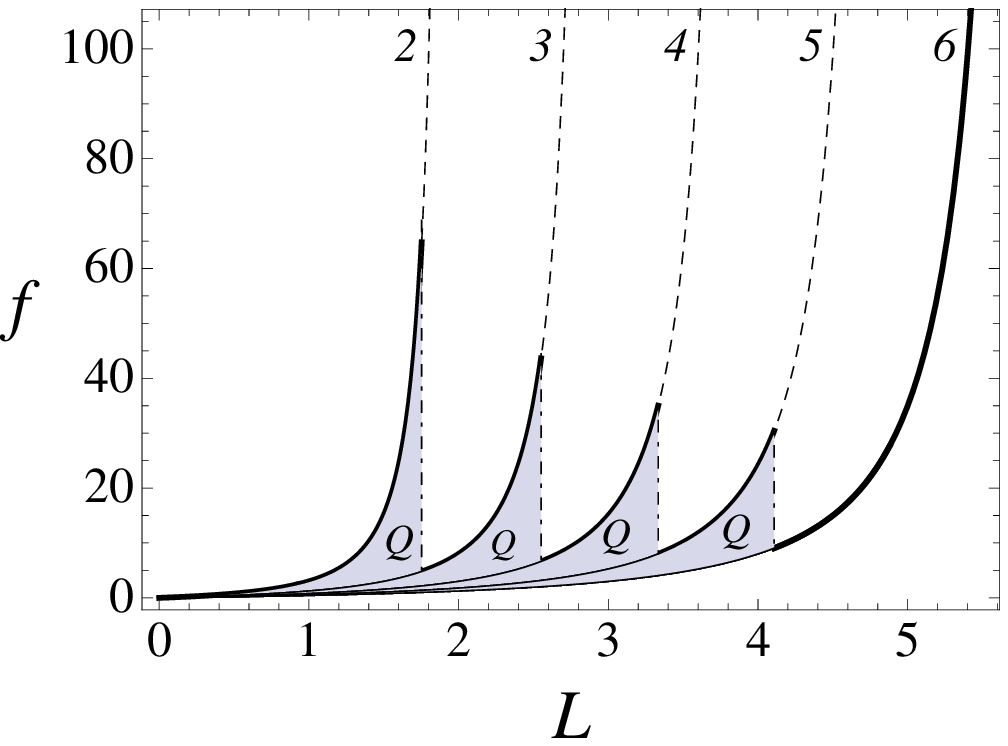}\\
a) & b) &c)\end{array}$$
\caption{Unfolding behavior for a system with $n_o=2$ and $n_t=6$. Here  we considered the parameters $Q = 10$, $\zeta = 0.1$ and $l_c = 1$. Each equilibrium path is labelled by the number $n$ of unfolded domains. \label{modelsoft}
}
\end{center}
\end{figure}

Under this global minimization hypothesis, one can show (see Fig. \ref{modelsoft}$_a$ and the proof in \cite{DMPS}) that the chain unfolds  following a sequence of single domain transitions.  Thus, the minimization of the total energy requires that the unfolding length  of the $n$-th domain $L^{\mbox{\tiny {\it un}}}= L^{\mbox{\tiny {\it un}}}(n)$ is obtained as the solution of the condition
$
\Phi(L, n) - \Phi(L, n+1)= 0, \hspace{0.6 cm} n=n_o,...,n_t-1$. This gives\begin{equation} L^{\mbox{\tiny {\it un}}}= L^{\mbox{\tiny {\it un}}}(n)=\frac{2
   n +1-\sqrt{4
   \zeta  n  (n +1)+1}}{2(1- \zeta) } \, l_c.
\label{lthr}\end{equation}

Here \begin{equation} \zeta= \frac{\kappa\, l_c}{Q}, \label{zetazeta} \end{equation}
is the main non-dimensional parameter of the system that measures the ratio between unfolding and elastic energy of the single hard domain. Observe that $L^{\mbox{\tiny {\it un}}}\in (0, L_c)$ and that $L^{\mbox{\tiny {\it un}}}$ increase with $n$
 so that the $n$-th branch delivers the global minimum of the energy in the interval
$$ L \in (L^{\mbox{\tiny {\it un}}}(n-1),L^{\mbox{\tiny {\it un}}}(n)), \,\,\,\,\, n\in(1,n_t-1).$$

Using (\ref{WLC}), we get the corresponding unfolding force 
\begin{equation}
f^{\mbox{\tiny {\it un}}}= f^{\mbox{\tiny {\it un}}}(n)=\frac{k_b T}{4 L_p}\frac{ 2 \zeta  n 
   (n +2)+1+(2 \zeta  n +1)
   \sqrt{4 \zeta  n  (n
   +1)+1}}{2 \zeta ^2
   n ^2}\label{fthr}.
\end{equation}

In Fig.\ref{modelsoft} we describe the unfolding curve of the chain deduced above. In the figure we schematically show the decomposition of the external work into unfolding (dissipated) and entropic energy. Observe that, according with the considered Maxwell hypothesis, the system follows an equilibrium branch until the elastic energy difference (between the two branches) equals the dissipation $Q$. Thus the chain follows the typical sawtooth transition path observed in  protein macromolecules with periodic transitions corresponding to the unfolding of single domains signaled by  the force drops (entropy discontinuities). 

Regarding the behavior of the system under unloading, we remark that a simplifying hypothesis of our model is {\it unfolding irreversibility}.   As a result if the system is unloaded it follows at fixed $n$ the same equilibrium branch until both stretch and force goes to zero. Under reloading the system follows again the same branch undergoing another hard-soft transition at  $L^{\mbox{\tiny {\it un}}}=L^{\mbox{\tiny {\it un}}}(n)$. Thus the unloading-reloading unfolding length is the same of the monotonically loaded system and the memory of the system is  restricted to the only maximum value attained in the past by the end-to-end length $L^{\mbox{\tiny {\it max}}}$.

More general assumptions in this direction can be considered by introducing an healing effect (see \cite{DMPZ}, \cite{Rief5}).

\vskip0.5 cm

\noindent {\it Remark 1.} We remark that our approach is general and not restricted to the chosen expression for the  energy (\ref{Phie}).  The choice here performed is dictated by the mathematical feasibility and by the ability of considering non-Gaussians (with variable limit chain extensibility) effects. Interesting possible extensions regard the possibility of considering variable unfolding energies \cite{DMPS} or  rate-dependent unfolding effects \cite{Qi}.

\section{Continuum Limit}\label{cl}

In this section, aimed to the deduction of a continuum three-dimensional model for macromolecular materials (see {\it e.g.} \cite{DeTommasi-Puglisi-Saccomandi2}), we analyze the continuum approximation of the macromolecule behavior described in the previous section. To get this limit we may formally  fix the {\it total unfolded length}, that by (\ref{lcc}) is given by 
\begin{equation}\label{tcl}  L_c^{\mbox{\tiny 1}}=n_t l_c,\end{equation}
 and consider the limit when both  $l_c\rightarrow 0$ and $n_t \rightarrow \infty$
   (a similar approach has been proposed  {\it e.g.} in \cite{PT} where the authors deduce a continuum damage model useful in the description of peeling in biological adhesion.
 A mathematical justification, based on a discrete-continuum $\Gamma$-limit has been delivered in  \cite{MPPT}). 
 
To this scope, we first introduce the  (continuum) {\it  internal variable}  \cite{Z}
\begin{equation}\nu:=\frac{n}{n_t}\in (\nu_o,1),\label{nu}\end{equation}
representing the unfolded fraction, with  $\nu= \nu_o=n_o/n_t$ in the initial configuration  and $\nu=1$ in the fully unfolded state. 
As a result the elastic energy $\Phi^e$ is formally given by the same expression (\ref{Phie}) 
 where  the total contour length now, by using (\ref{lcc}),  is given by
\begin{equation}\label{Lcnu} L_c= L_c(\nu) =\nu L_c^{\mbox{\tiny 1}}\in (L_c^{\mbox{\tiny 0}},L_c^{\mbox{\tiny 1}}).
\end{equation}
Thus in this case the energy  depends on the continuous internal variable $\nu$. In particular $L_c=L_c^{\mbox{\tiny 0}}$ in the initial configuration ($\nu=\nu_o$) and $L_c=L_c^{\mbox{\tiny 1}}$ in the unfolding saturation configuration ($\nu=1$). Moreover, in view of (\ref{Lcnu}), also the {\it relative stretch} $\lambda_{\mbox{\tiny {\it rel}}}$ of the chain depends, for the given end-to-end length $L$,  on the continuum unfolding  variable $\nu$ according with the following relation: 
\begin{equation}\lambda_{\mbox{\tiny {\it rel}}}=\lambda_{\mbox{\tiny {\it rel}}}(L,\nu)=\frac{L}{L_c(\nu)}=\frac{L}{\nu L_c^{\mbox{\tiny 1}}}.\label{etac}\end{equation}
The total energy $\Phi$ in (\ref{llll}) and (\ref{Phie})  can then be expressed as\begin{equation}
\label{Toten}
 \Phi=\Phi(L,\nu)=\Phi_e(L,\nu)+\nu n_t Q=
\kappa  \frac{\lambda_{\mbox{\tiny {\it rel}}} ^2}{1-\lambda_{\mbox{\tiny {\it rel}}} }  \nu L_c^{\mbox{\tiny 1}}+ \nu n_t Q.\end{equation}
Finally observe that the equilibrium equation 
\begin{equation}f= \partial_L \Phi_e(L,\nu)\label{eeqq}\end{equation} gives again the expression (\ref{WLC}) of the equilibrium force.

\begin{figure}[!h]\begin{center}
\includegraphics[scale=0.3]{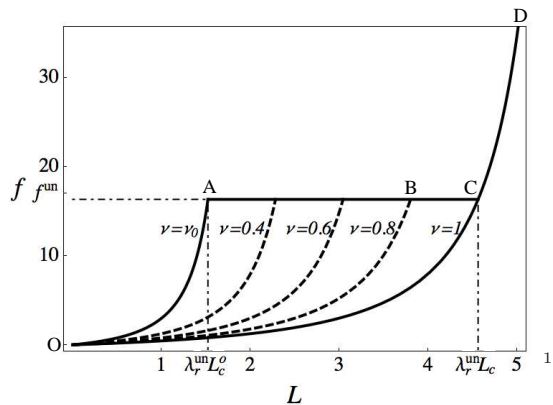} 
\vspace{-1.25 cm}

\hspace{6.75 cm} {\tiny 1}\vspace{0.6 cm}
\caption{Force-length relation for a continuum system with the same parameters of Fig.\ref{modelsoft}. Each equilibrium path is labelled by the unfolding fraction $\nu$. \label{modelcont}
}\end{center}
\end{figure}

An important result in the analysis of the continuum limit behavior (see Fig.\ref{modelcont}) of the unfolding chain is that  the unfolding events are characterized by a {\it fixed unfolding force} (stress plateau) obtained by  (\ref{fthr}) and (\ref{nu}) in the limit $n_t \rightarrow \infty$  
\begin{equation}f^{\mbox{\tiny {\it un}}}=\kappa\frac{2 \sqrt{\zeta} +1}{
   \zeta  }\end{equation}
\noindent and a {\it fixed unfolding relative stretch} $\lambda_{\mbox{\tiny {\it rel}}}=\lambda_{\mbox{\tiny {\it rel}}}^{\mbox{\tiny {\it un}}}$, that using (\ref{Lcnu}) and (\ref{nu}) in (\ref{lthr}) is given by 
\begin{equation} \lambda_{\mbox{\tiny {\it rel}}}^{\mbox{\tiny {\it un}}}=\frac{1}{\sqrt{\zeta}+1}.
\label{luuu}\end{equation}
 Using (\ref{etac}), we obtain in the continuum limit the unfolding length
$$L^{\mbox{\tiny {\it un}}} =\nu  \lambda_{\mbox{\tiny {\it rel}}}^{\mbox{\tiny {\it un}}} \,L_c^{\mbox{\tiny 1}}=\nu\frac{ L_c^{\mbox{\tiny 1}}}{\sqrt{\zeta}+1}\in (\lambda_{\mbox{\tiny {\it rel}}}^{\mbox{\tiny {\it un}}} \, L_c^{\mbox{\tiny 0}}
  ,\lambda_{\mbox{\tiny {\it rel}}}^{\mbox{\tiny {\it un}}}  \, L_c^{\mbox{\tiny 1}}
 ).$$

As a result the unfolding behavior of the chain is assigned by the parameters $P$, $l_c$ and $\zeta$ in (\ref{zetazeta})   (or analogously the unfolding stretch $\lambda_{\mbox{\tiny {\it rel}}}^{\mbox{\tiny {\it un}}}$ in (\ref{luuu})) that can be deduced by the analysis of the unfolding behavior of the  single folded domain that fixes the unfolding force and the relative stretch. Instead, the  initial number $n_o$ and the total number of elements $n_t$   (or analogously $L_c^{\mbox{\tiny 0}}$ and $L_c^{\mbox{\tiny 1}}$) 
 assign  the initial elastic limit  $L=\lambda_{\mbox{\tiny {\it rel}}}^{\mbox{\tiny {\it un}}} \,L^o_c$ of the end-to-end length and the unfolding saturation limit $L=\lambda_{\mbox{\tiny {\it rel}}}^{\mbox{\tiny {\it un}}} \,L^{\mbox{\tiny 1}}_c$ (see again Fig.\ref{modelcont}).

The behavior of the system is illustrated in Fig.\ref{modelcont}. Thus suppose that we assign an increasing end-to-end length $L$, 
the unravelling  begins at  $L=\lambda_{\mbox{\tiny {\it rel}}}^{\mbox{\tiny {\it un}}}\, L_c^{\mbox{\tiny 0}}$ (point A in the figure) that represents the initial elastic threshold. If $L$  is increased further, the chain unfolds following the stress plateaux (A-B-C in the figure) at $f=f^{\mbox{\tiny {\it un}}}$.  The unfolding saturation effect is attained at
$L=\lambda_{\mbox{\tiny {\it rel}}}^{\mbox{\tiny {\it un}}} \,L_c^{\mbox{\tiny 1}}$ (point C) 
so that for $L^{\mbox{\tiny {\it max}}}\geq \lambda_{\mbox{\tiny {\it rel}}}^{\mbox{\tiny {\it un}}} \,L_c^{\mbox{\tiny 1}}$ the behavior is again elastic with fixed  $\nu=1$ (curve O-C-D). 

Observe that if we unload before unfolding saturation is attained ({\it e.g.} point B in the figure), the system follows an equilibrium branch at fixed  $\nu$ (curve B-O). Finally, if we reload again, the system follows the same branch until $L=L^{\mbox{\tiny {\it max}}}$ (B) when the unfolding fraction $\nu$ starts increasing again (path B-C). 

Notice that, as in the discrete case considered in the previous section, the memory of the system is restricted to $L^{\mbox{\tiny {\it max}}}$. In particular we have
\begin{equation} \label{nuL}\nu=\bar \nu (L^{\mbox{\tiny {\it max}}})=\frac{L^{\mbox{\tiny {\it max}}}}{\lambda_{\mbox{\tiny {\it rel}}}^{\mbox{\tiny {\it un}}}\, L_c^{\mbox{\tiny 1}}}=\frac{L^{\mbox{\tiny {\it max}}}} {L_c^{\mbox{\tiny 1}}}(\sqrt{\zeta}+1) \hspace{ 1 cm} L^{\mbox{\tiny {\it max}}} \in (\lambda_{\mbox{\tiny {\it rel}}}^{\mbox{\tiny {\it un}}}\,L_c^{\mbox{\tiny 0}}
  ,\lambda_{\mbox{\tiny {\it rel}}}^{\mbox{\tiny {\it un}}}\,  L_c^{\mbox{\tiny 1}}
 ).\end{equation}
\vspace{0.3 cm}

\subsection{Thermodynamical consistence}

The continuous model previously deduced can be inscribed in the framework of Thermodynamics with internal state variables \cite{CG} in the special simple setting of {\it isothermal processes}. 
In particular, the state of the chain is assigned by  a single `external variable' $L$ and a single `internal variable' $\nu$. Following \cite{CG},  we may  introduce the rate of energy dissipation $\Gamma$
$$\Gamma=-\dot \Phi_e +f \dot L$$
and write the classical (Clausius-Duhem) dissipation inequality \begin{equation}\Gamma= \left ( -\partial_L  \Phi_e(L,\nu) +f \right ) \dot L+g  \dot \nu\geq 0,  \label{dis} \end{equation}
ensuring the thermodynamical consistence of the model.
Here
\begin{equation} \label{ddff}g=g(L,\nu)=-\partial_{\nu}\Phi_e(L, \nu)=\kappa\frac{  \lambda_{\mbox{\tiny {\it rel}}} ^2}{(1-\lambda_{\mbox{\tiny {\it rel}}})^2} L_c^{\mbox{\tiny 1}}> 0\end{equation}
represents the {\it driving force} working for {\it the chain length growth} due to the increase of the unfolding fraction \cite{Z}.

Using the equilibrium equation (\ref{eeqq}),  (\ref{dis}) reduces to the  dissipation inequality
\begin{equation}\Gamma=g  \dot \nu\geq 0 \label{rCD},\end{equation}
that in view of  (\ref{ddff}) shows that the Clausius-Duhem inequality is respected if an only if  our irreversibility hypothesis
$$\dot \nu \geq 0$$
is fulfilled.

To show that our model is consistent with the classical framework of Continuum Thermodynamics  with internal state variables (\cite{GNS}), we observe that the unfolding threshold  can be equivalently assigned by one of the three following conditions:
\begin{equation}\left \{ \begin{array} {l}h_L(L)=L-L^{\mbox{\tiny {\it un}}}(\nu)=0, \vspace{0.5 cm} \\ h_f(f)=f-f^{\mbox{\tiny {\it un}}}=0, \vspace{0.5 cm} \\ h_{\lambda_{\mbox{\tiny {\it rel}}}}(\lambda_{\mbox{\tiny {\it rel}}})=\lambda_{\mbox{\tiny {\it rel}}}-\lambda_{\mbox{\tiny {\it rel}}}^{\mbox{\tiny {\it un}}}=0, \end{array} \right . \label{thresh}\end{equation}
defining in particular the present elastic domain as a function of $\nu$. 
We may then deduce the flow rule 
$$\dot \nu=\dot \nu(L,\dot L, \nu)=\left \{ \begin{array}{llll} 0 & \mbox{ if } &  L< L^{\mbox{\tiny {\it un}}}, & \mbox{ elastic regime,} \vspace{0.2 cm}\\
0 & \mbox{ if } &  L= L^{\mbox{\tiny {\it un}}}, \dot L\leq 0, & \mbox{ elastic unloading,} \vspace{0.2 cm}\\
\displaystyle \frac{\dot L}{\lambda_{\mbox{\tiny {\it rel}}}^{\mbox{\tiny {\it un}}} \,L_c^{\mbox{\tiny 1}}} ,
 & \mbox{ if } &   L= L^{\mbox{\tiny {\it un}}}, \dot L>0, & \mbox{ unfolding regime.} \\ \end{array} \right .
$$
Moreover, to evaluate the {\it dissipation potential} $D$ we observe that  during unravelling, using (\ref{ddff}), (\ref{luuu}) and (\ref{zetazeta})  we have
$$g=g(L^{\mbox{\tiny {\it un}}}(\nu),\nu)=n_t \, Q$$ so that 
$$D=D(\dot \nu)=n_t \, Q \,  \dot \nu.$$
In particular, we obtain  a constant dissipation rate $\Gamma$ and a rate-independent dissipation behavior (the potential is a homogeneous function of degree one).
\vspace{0.13 cm}

Finally, it is interesting to observe, both from a theoretical and numerical point of view, that the 
unfolding behavior of the system can be obtained through a {\it variational approach, considering a Griffith-type minimization scheme for the total} ({\it elastic plus unfolding}) {\it   potential energy}
 $$G(L, \nu)=\Phi(L,\nu)- f L.$$ 

Indeed if we consider  the constrained minimization problem
$$\min_{\nu \geq \bar \nu, L} G(L, \bar \nu),$$
where $\bar \nu$ is the present value of unfolded fraction,
the  evolution equations result  as classical Kuhn-Tucker minimization conditions (see {\it e.g.}  \cite{Ba}) for the Lagrangian function
${\cal L}= G(L,\bar \nu)+ \mu\, (\bar \nu-\nu)$:
 \begin{equation}\left \{ \begin{array}{l}   \displaystyle
\partial_L {\cal L}=L_c^{\mbox{\tiny 1}} \partial_L \varphi_e -f=0,\vspace{0.2 cm}\\ \displaystyle \partial_{\nu} {\cal L}=n_t Q-g-\mu=0, \vspace{0.2 cm}\\ \displaystyle
\nu \geq \bar \nu,\vspace{0.2 cm}\\ \displaystyle
\mu \geq 0, \vspace{0.2 cm}\\  \displaystyle
\mu (\nu -\bar \nu) = 0.
\end{array} \right .
\label{KT}\end{equation}

By (\ref{KT})$_1$ we deduce  the equilibrium equation
$$f= \partial_L \Phi_e;$$

\noindent by (\ref{KT})$_{2,5}$ we obtain the {\it consistency condition}
$$(n_t Q-g) (\bar \nu-\nu)= 0,$$

\noindent ensuring that the unfolding can happen only on the thresholds defined in (\ref{thresh}). Moreover, since $\mu \geq 0$, (\ref{KT})$_{2}$ gives the {\it admissibility condition} (see (\ref{thresh})) 
$$g\leq n_t Q \Rightarrow f\leq f^{\mbox{\tiny {\it un}}},  L\leq L^{\mbox{\tiny {\it un}}}.$$\vspace{0.3 cm}

\noindent {\it Remark 2}.\hspace{0.2 cm} We may observe that since the energy is positive semidefinite these conditions are also sufficient to attain the global minimum of the total energy $G$ (see {\it e.g.} \cite{Ba}), representing the {\it unique}  solution due to the convexity of the energy function. \vspace{0.5 cm}

\noindent {\it Remark 3}.\hspace{0.2 cm} It is easy to verify that the conditions (\ref{KT}) correspond to the classical condition of {\it maximal dissipation} adopted in non-equilibrium Thermodynamics \cite{Z}. Indeed, 
if we consider the incremental variational problem for (\ref{rCD})
$$\displaystyle \max_{g- n_t Q\leq 0} \Gamma=\max_{g- n_t Q \leq 0}  g \dot \nu$$
we obtaion the Kuhn-Tucker condition
$\displaystyle (g-n_t Q)\dot \nu=0$.\vspace{0.5 cm}

\subsection{Total, elastic and residual stretches}\label{tttfff}

Based on a classical Statistical Mechanics approach  \cite{Rubinstein}, here we introduce the notion of elastic, residual and total stretch for the macromolecule, fundamental in the following deduction of the three-dimensional constitutive equation.

First we observe that  the {\it natural length}  (zero force)  $L_p$ of the entropic chain  can be expressed, according to a know result reported {\it e.g.} in \cite{Rubinstein}, as follows:
\begin{equation}L_p=\sqrt{\bar n} b,\end{equation}
where $b$ is the length of the Kuhn segments and $\bar n$  the number of Kuhn segments in the present configuration. Since
$$\bar n=n_k n,$$ where $n_k=l_c/b$ is the number of Kuhn segments inside each unfolded domain, we obtain 
\begin{equation}L_p=\sqrt{\nu \bar n_{t}} b,\label{llpp}\end{equation}
showing the explicit dependence of the natural length on the percentage $\nu$ of unfolded domains.
Here\begin{equation}\label{nnt}\bar n_t=\frac{l_c}{b}n_t=n_k n_t\end{equation}
is the total number of Kuhn segments.

\begin{figure}[!h]
\begin{center}\includegraphics[scale=0.32]{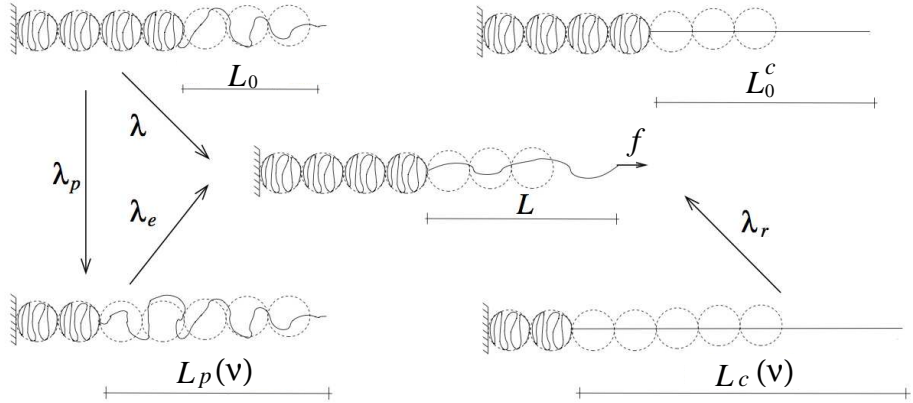} 
\caption{Scheme of the elastic, permanent, total and relative stretches.
} \label{scca}
\end{center}
\end{figure}

Thus, if we denote by
\begin{equation}\label{Lo}L_o=\sqrt{\nu_o \bar n_{t}} b\end{equation}
the initial natural end-to-end length,
we may define the following stretch measures  (see  Fig.\ref{scca})
\begin{equation} \left \{ \begin{array}{ll}\displaystyle
\lambda=\frac{L}{L_o}, & (total)\,\,\, stretch, \vspace{0.2 cm}\\ \displaystyle
\lambda_e=\frac{L}{L_p(\nu)}, & elastic\,\,\, stretch,  \vspace{0.2 cm}\\ \displaystyle
\lambda_p=\frac{L_p(\nu)}{L_o}, & permanent\,\,\, stretch,
\end{array} \right . \label{str} \end{equation}

\noindent verifying the classical deformation composition $\lambda=\lambda_e \lambda_p$. Moreover it is easy to verify (see (\ref{etac}) and (\ref{nnt})) that
\begin{equation}\label{abc}\lambda=\nu \sqrt{\frac{\bar n_t}{\nu_o }}\,\,\lambda_{\mbox{\tiny {\it rel}}}, \,\,\,\,\,
\lambda_e=\sqrt {\nu \bar n_t } \, \lambda_{\mbox{\tiny {\it rel}}}, \,\,\,\,\,\lambda_p(\nu)=\sqrt{\frac{\nu}{\nu_o}}.\end{equation}

For the following three-dimensional extension we also rephrase the unravelling evolution law (\ref{nuL}) in terms of the stretch $\lambda$. Since according with (\ref{str})$_1$ the maximum attained stretch  $\lambda^{\mbox{\tiny {\it max}}}$ corresponds to $L^{\mbox{\tiny {\it max}}}$, using (\ref{nuL}), (\ref{tcl}) and (\ref{Lo}) we obtain
\begin{equation} \nu= \sqrt{\frac{\nu_o}{\bar n_t}}\,\, \frac{\lambda^{\mbox{\tiny {\it max}}}}{\lambda_{\mbox{\tiny {\it rel}}}^{\mbox{\tiny {\it un}}}},   \mbox{ with }\lambda^{\mbox{\tiny {\it max}}} \in (\lambda_o,\lambda_1)\label{nunu}.\end{equation}

\noindent Here 
\begin{equation}
\lambda_o=\sqrt{\nu_o \bar n_t} \lambda_{\mbox{\tiny {\it rel}}}^{\mbox{\tiny {\it un}}}, \hspace{1 cm} \bigskip \lambda_1=\sqrt{\frac{ \bar n_t}{\nu_o  }}  \lambda_{\mbox{\tiny {\it rel}}}^{\mbox{\tiny {\it un}}} \label{thress}
\end{equation}
 are the stretches  corresponding to the initial unfolding threshold (obtained by (\ref{abc}) with $\nu=\nu_o$ and $ \lambda_{\mbox{\tiny {\it rel}}}= \lambda_{\mbox{\tiny {\it rel}}}^{\mbox{\tiny {\it un}}}$) and to the attainment of the totally unfolded configuration (obtained by (\ref{abc}) with $\nu=1$ and $ \lambda_{\mbox{\tiny {\it rel}}}= \lambda_{\mbox{\tiny {\it rel}}}^{\mbox{\tiny {\it un}}}$), respectively. 
 
 Finally we introduce the limit stretch $\lambda_c$ corresponding to the attainment of the contour length (when $L=L_c$ or equivalently $\lambda_{\mbox{\tiny {\it rel}}}=1$) that, using (\ref{abc})$_1$ and (\ref{nunu}) is given by 
   \begin{equation}\lambda_c=\nu \sqrt{\frac{\bar n_t}{\nu_o }}=\frac{\lambda^{\mbox{\tiny {\it max}}}}{\lambda_{\mbox{\tiny {\it rel}}}^{\mbox{\tiny {\it un}}}}.\label{lambdac} \end{equation}
 Observe that   \begin{equation}\lambda_{\mbox{\tiny {\it rel}}}=\frac{\lambda}{\lambda_c}.\label{ttt} \end{equation}

\section{Three-dimensional constitutive model}\label{3D}

In this section we consider the second scale change that let us deduce a  three-dimensional  {\it macroscopic} model based on the  behavior of the single macromolecule obtained in the previous section. As a specific effect, due to the variation of the natural length of the chains, we obtain the insurgence of volume variations and the consequent effect of residual stresses and stretches. The resulting continuum model can be inscribed within the field of growth non-linear mechanics (see \cite{DQ}, \cite{AG}, \cite{GON} and reference therein).\vspace{0.2 cm}

To connect the macroscopic stretches to the chains elongations, different possible Cauchy-Born type schemes can be considered (see {\it e.g.} \cite{Er}). A typical approach for macromolecular materials, known as {\it affine hypothesis} \cite{Rubinstein}, identify macroscopic stretches with chains elongations. In this perspective several microstructure based schemes have been considered by introducing different chain cell structures ({\it e.g.}  three, four and 8-chain model, see \cite{BA} and \cite{Beatty}) with a similar assumption of coincidence of cell and macroscopic stretches. More refined schemes consider a continuum distribution of  chains such  as in the full network model (see  \cite{WV} and \cite{MGL}). In this last case the energy density depends on the orientation and the spatial distribution of the chains (the resulting model can be numerically heavy: see \cite{BO} for an efficient numerical approach to this problem). Also, non-affine effects can be considered as shown in the paper \cite{CMD}: the authors argue, by considering a phantom network model, that these non-affine effects influence only the shear modulus of the network whereas it does not modify the functional form of the macroscopic constitutive response function. 

All these schemes can be considered for a micro-macro scale change of our model. Here, aimed at a clear analytical description, we follow the classical and efficient Arruda-Boyce 8-chain approach
 \cite{AB}    (see also the extension of this model proposed in \cite{K} focused on non-affine effects). Other possible assumptions will be the subject of our future work. \vspace{0.2 cm}

Let $\mbox{\boldmath{$f$}}$ be the (macroscopic)  deformation function which carries a continuous body (representing a three-dimensional network of unfolding macromolecules) from the reference configuration ${\cal B}_o$ to the present configuration $\cal B$. Let  then $\mbox{\boldmath{$B$}}$ be the left Cauchy-Green strain tensor
\begin{equation} \mbox{\boldmath{$B$}}=\nabla \mbox{\boldmath{$f$}}( \nabla\mbox{\boldmath{$f$}})^T=\lambda^2_1 \mbox{\boldmath{$e$}}_1 \otimes  \mbox{\boldmath{$e$}}_1+\lambda^2_2 \mbox{\boldmath{$e$}}_2 \otimes  \mbox{\boldmath{$e$}}_2+\lambda^2_3\mbox{\boldmath{$e$}}_3 \otimes  \mbox{\boldmath{$e$}}_3,\label{CG}\end{equation}
where $\lambda_i$ and $\mbox{\boldmath{$e$}}_i$, $i=1,2,3$, are the principal stretches and the principal unit vectors, respectively. 
In the  Arruda-Boyce 8-chain model the  chains are distributed along  the eight diagonals connecting the vertexes to the  center of a unit cube with  
 faces normal to  $\mbox{\boldmath{$e$}}_1, \mbox{\boldmath{$e$}}_2,\mbox{\boldmath{$e$}}_3$  (see (\ref{CG})). This cube cell is assumed to undergo the same stretch measured at the macroscopic scale.  As shown in   \cite{AB}, this cell  is energetically equivalent to a cubic cell with all chains aligned in the direction $$\mbox{\boldmath{$m$}}=\frac{1}{\sqrt{3}}(\mbox{\boldmath{$e$}}_1+\mbox{\boldmath{$e$}}_2+\mbox{\boldmath{$e$}}_3).$$ These chains are characterized by the  stretch
\begin{equation}\lambda=\sqrt{\frac{I}{3}}
\label{princ} \end{equation}
where $I= \mbox{tr \boldmath{$B$}}$.
Since, as shown  by Kearsley \cite{Kearsley}, the first invariant $I$ is equal to ``{\it three times the square of the stretch ratio of an infinitesimal line element averaged over all possible orientations}'', the 8-chain approximation let us deduce the evolution of the chain unfolding fraction and residual stretch  in an ``averaged sense" with all chains undergoing the same elongation, the same unfolding, and the same residual stretch. 
 \begin{figure}[!h]
\begin{center}\includegraphics[scale=0.32]{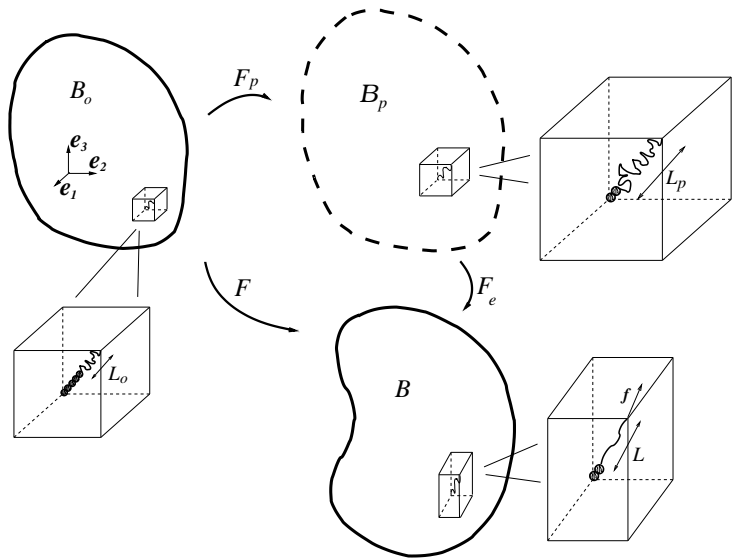} 
\caption{Scheme of the reference, relaxed and deformed configuration in the three dimensional setting.
} \label{scc}
\end{center}
\end{figure}

 Due to the   the existence (see Sect. \ref{cl}) of a variable natural configuration with a consequent  insurgence of  residual stretches induced by  unfolding, we inscribe our model in the  framework of Growth Mechanics. We thus consider the classical multiplicative decomposition (proposed in \cite{Rodr}, see the scheme in Fig.\ref{scc}) of the strain tensor   \begin{equation}\mbox{\boldmath{$F$}}=\mbox{\boldmath{$F$}}_e\mbox{\boldmath{$F$}}_p.\label{FF}\end{equation} 
Here $\mbox{\boldmath{$F$}}_p$ is the {\it growth tensor}, carrying the reference configuration ${\cal B}_o$ to the  {\it relaxed configuration} ${\cal B}_p$, representing the unstressed configuration. This configuration in general cannot be described by a compatible deformation \cite{DQ}. Instead, the {\it elastic tensor} $\mbox{\boldmath{$F$}}_e$ represents the deformation measured starting from the relaxed configuration ${\cal B}_p$. 
 In the following, as usual for many biological and polymeric materials,  we assume {\it elastic incompressibility}, so that
\begin{equation}\label{kkk}\det \mbox{\boldmath{$F$}}_e=1.\end{equation}

 Coherently with the 8-chain assumption of homogeneous chain stretch, we here assume that the growth tensor  is spherical. Thus, by (\ref{str})$_3$, (\ref{llpp}) and (\ref{Lo}), it assumes the form \begin{equation}\label{Fp}\mbox{\boldmath{$F$}}_p=\sqrt{\frac{\nu}{\nu_o}} \mbox{\boldmath{$I$}}.\end{equation}
 As a result. we consider a continuum with a volumetric growth  (see \cite{Rodr} and \cite{EM})  induced by unfolding. It is important to remark that in our model the evolution of the growth tensor is a result of our minimization approach, because it explicitly depends   on the microstructure parameter $\nu$.
We remark that more general assumptions with anisotropic affects can be considered \cite{HOS} and that in particular the isotropic unfolding hypothesis (see the discussion in the following section) can induce an underestimation of the unfolding effect and of the resulting softening and residual stretch.\vspace{0.2 cm}

Based on these assumptions, we here deduce the expression of the macroscopic energy density. First, we observe that the first invariant $I^e$ of the elastic 
right Cauchy-Green strain tensor $ \mbox{\boldmath{$B$}}_e=\mbox{\boldmath{$F$}}_e{\mbox{\boldmath{$F$}}_e}^T$ is related to $I$ by 

\begin{equation}I=\frac{\nu}{\nu_0} I^e.
\label{III} \end{equation}
Then, we may relate the chain relative stretch $\lambda_{\mbox{\tiny {\it rel}}}$ to the 
first invariants $I$ and $I^e$. Indeed, using
 (\ref{princ}), (\ref{lambdac}) and (\ref{ttt}) we have that 
\begin{equation}\label{dom} 
\lambda_{\mbox{\tiny {\it rel}}}=\sqrt{\frac{I}{I_c}}=\sqrt{\frac{I^e}{I^e_c}}\end{equation}
\noindent where we introduced the critical value of the  
 the first invariant 
 \begin{equation}\label{icic} I_c=I_c(\nu)=3 \lambda_c^2=3\nu^2 \frac{\bar n_t}{\nu_o }, \hspace{1 cm} I_c\in(I^o_c,I^1_c),
\end{equation} 
corresponding in the 8-chain model to the limit condition of the attainment of the   contour length. Using (\ref{III}) the corresponding limit value $I^e_c$ of $I^e$ is given by 
\begin{equation} I^e_c=I^e_c(\nu)=\frac{\nu_0}{\nu} \label{Ic}I_c.\end{equation} 
In (\ref{icic}) 
\begin{equation}I_c^o=I_c(\nu_o)=3 \nu_o \bar n_t \,\,\, \mbox{ and } \,\,\,I_c^1=I_c(1)=3\frac{ \bar n_t} {\nu_o}\label{Ioc}\end{equation}
represent the virgin  and saturations unfolding thresholds, respectively. 
Finally, by (\ref{icic}) and (\ref{Ioc}), we have
\begin{equation}\frac{\nu}{\nu_0}=\sqrt{\frac{I_c}{I^o_c}}.\label{niIc}\end{equation}
Based on this relation, in the continuum three-dimensional setting we choose $I_c$ (or equivalently $I_c^e$) as internal state variable. As in (\ref{nunu}) we may relate the internal variable $I_c$ to $I^{\mbox{\tiny {\it max}}}$, the maximum previously attained value of $I$. Indeed, since   $I^{\mbox{\tiny {\it max}}}=3( \lambda^{\mbox{\tiny {\it max}}})^2$, where here $\lambda^{\mbox{\tiny {\it max}}}$ is the maximum attained stretch of the representative chain, using (\ref{princ}) and (\ref{lambdac})$_2$ we obtain
\begin{equation}\label{IcIc}
I_c=\frac{I^{\mbox{\tiny {\it max}}}}{(\lambda_{\mbox{\tiny {\it rel}}}^{\mbox{\tiny {\it un}}}\,)^2}, \hspace{1 cm} I^{\mbox{\tiny {\it max}}}\in(I^o,I^1),
\end{equation}
with $\lambda_{\mbox{\tiny {\it rel}}}^{\mbox{\tiny {\it un}}}$ the material parameter in (\ref{luuu}) . Here (see Eq. (\ref{thress}))
$$I^o=3 \nu_o \bar n_t (\lambda_{\mbox{\tiny {\it rel}}}^{\mbox{\tiny {\it un}}}\,)^2\,\,\, \mbox{ and }  \,\,\,I^1=3\frac{ \bar n_t} {\nu_o}(\lambda_{\mbox{\tiny {\it rel}}}^{\mbox{\tiny {\it un}}}\,)^2$$
are the virgin and saturations unfolding thresholds of the invariant $I$, respectively.

We are now in position to deduce the macroscopic energy density, starting from  the energy of the representative chain (\ref{Toten}). First we introduce the number $N$ of chains per unit volume of the relaxed (unfolded) configuration ${\cal B}_p$. Observe that this corresponds to a variable  number $\displaystyle N_o=N/\det \mbox{\boldmath{$F$}}_p$ of chains  per unit volume of the reference configuration ${\cal B}_o$; this is a consequence of the unfolding phenomenon which changes the length of the chains. Then, based on an entropy additivity assumption,  the three dimensional energy density is given by  \begin{equation}\varphi_e=N \Phi_e(L,\nu).\label{defdef}\end{equation}

Using   (\ref{Phie}) and (\ref{abc})  we obtain 
\begin{equation}\varphi_e= N \kappa   \frac{\lambda_{\mbox{\tiny {\it rel}}}^2}{1-\lambda_{\mbox{\tiny {\it rel}}}} \nu L_c^{\mbox{\tiny 1}}-c, \label{fifi}
\end{equation}
where $c$ is the usual normalizing constant such that $\varphi_e=0$ in the reference configuration. 
As a result, since $L_c^{\mbox{\tiny 1}}=\bar n_t b$, we obtain, by using (\ref{dom}) and  (\ref{Ic}),

\begin{equation}\varphi_e= \frac{\mu_o}{2}\frac{I^e }{1-\sqrt{\frac{ I^e}{I^e_c}}}-c,\label{contphi}
\end{equation}
where
\begin{equation}\mu_o= \frac{2}{3} \kappa N b=\frac{N   b }{6 p} k_B T \label{muo}\end{equation}
is the  (virgin) {\it  infinitesimal shear modulus} and $c=\frac{3 \mu_o}{2\left(1-\sqrt{\frac{3}{I^e_c}}\right )}$. 

As a result, using  (\ref{FF}), (\ref{Ic}) and (\ref{niIc}), we finally obtain  the macroscopic constitutive law, giving the Cauchy stress tensor
\begin{equation}\mbox{\boldmath{$T$}}=-p \mbox{\boldmath{$I$}}+\partial_{{\mbox{\boldmath{\tiny $F$}}}_e}\varphi(I^e,I^e_{c}) \mbox{\boldmath{$F$}}_e^T=-p \mbox{\boldmath{$I$}}+
\aleph (I,I_{c})
\mbox{\boldmath{$B$}}, \label{equilb} \end{equation}

\noindent where  the response coefficient $\aleph$ is given by $$\displaystyle   \aleph (I,I_c)
=\frac{\mu_o}{2} \frac{2-\sqrt{ \frac{I}{I_c}}}{\left (1-\sqrt{\frac{I}{ I_c}}\right )^2}\left (\frac{I_c^o}{I_c}\right )^{\frac{1}{2}},$$
and  $p$ is the pressure arising from the kinematic constrain (\ref{kkk}).

We remark that in our model the microstructure limit threshold $I_c$ (or equivalently $I^e_c$) takes the same role of the limit threshold of the classical Gent model in rubber elasticity (see {\it e.g.} \cite{H7} and \cite{PugSac}).
Here $I_c$ can
be seen as a fair (and natural) measure of the average contour length of the
chains composing the  network. As a consequence in the following, based on (\ref{niIc}), we consider $I_c$ as a natural macroscopic measure of the unravelling phenomenon; in particular observe that it defines the residual stretch tensor, because  by see (\ref{Fp}) and (\ref{niIc}) we have
 \begin{equation}\mbox{\boldmath{$F$}}_p=\left(\frac{I_c}{I_c^o}\right)^{\frac{1}{4} }\mbox{\boldmath{$I$}}.\label{FpFp}\end{equation}
 Also we observe that  in view of  (\ref{niIc})  our irreversibility assumption $\dot \nu\geq 0$ is equivalent to the condition $\dot I_c\geq 0$.\vspace{0.2 cm}

In terms of the Piola stress the constitutive Eq. (\ref{equilb}) takes the form
\begin{equation}\mbox{\boldmath{$S$}}=-p \left (\frac{I_c^o}{I_c}\right )^{\frac{3}{4}}\mbox{\boldmath{$F$}}^{-T}+
\aleph (I,I_{c})\left (\frac{I_c^o}{I_c}\right )^{\frac{3}{4}}
\mbox{\boldmath{$F$}}. \label{piola} \end{equation}

In particular we observe that approaching the primary loading path $I=I^{\mbox{\tiny {\it max}}}$, using (\ref{IcIc}), we can evaluate the value $ \bar \aleph (I)$ of $\aleph$ for the generic equilibrium branch:  
\begin{equation}\displaystyle    \bar \aleph (I)=\displaystyle    \aleph (I,(\lambda_{\mbox{\tiny {\it rel}}}^{\mbox{\tiny {\it un}}})^2 I)
=\frac{\mu_o }{2} \frac{2 \lambda_{\mbox{\tiny {\it rel}}}^{\mbox{\tiny {\it un}}}-(\lambda_{\mbox{\tiny {\it rel}}}^{\mbox{\tiny {\it un}}}\,)^2}{\left (1-\lambda_{\mbox{\tiny {\it rel}}}^{\mbox{\tiny {\it un}}}\, \right )^2} \left (\frac{I_c^o}{I}\right )^{\frac{1}{2}}.\label{Hbar}\end{equation}

  We may summarize the previous behavior as follows:\vspace{0.2 cm}
  
  \begin{equation} \mbox{\boldmath{$T$}}\!\! =\!\!\left \{ \begin{array}{llll} \!\!\!\!-p \mbox{\boldmath{$I$}}+
\aleph (I,I_c^o)
\mbox{\boldmath{$B$}}& \!\!\mbox{if} \!\!& I^{\mbox{\tiny {\it max}}}\leq I_o & \mbox{virgin elastic,} \vspace{0.2 cm}\\ 
\!\!\!\!-p \mbox{\boldmath{$I$}}+
\aleph (I,\frac{I^{\mbox{\tiny {\it max}}}}{(\lambda_{\mbox{\tiny {\it rel}}}^{\mbox{\tiny {\it un}}}\,)^2})
\mbox{\boldmath{$B$}} &\!\! \mbox{if} \!\!& I_o< I^{\mbox{\tiny {\it max}}} <I_1,  \dot I^{\mbox{\tiny {\it max}}}=0 & \mbox{fixed unfolding fraction,} \vspace{0.2 cm}\\ 
\displaystyle 
\!\!\!\! -p \mbox{\boldmath{$I$}}+\bar \aleph (I) \mbox{\boldmath{$B$}}
 &\!\! \mbox{if}\!\! &I_o< I^{\mbox{\tiny {\it max}}} <I_1, \dot I^{\mbox{\tiny {\it max}}}>0 & \mbox{growing unfolding fraction,}\vspace{0.2 cm} \\\!\!\!\! -p \mbox{\boldmath{$I$}}+
\aleph (I,I_c^1)
\mbox{\boldmath{$B$}} &\!\!  \mbox{if} \!\!& I^{\mbox{\tiny {\it max}}}\geq I_1 & \mbox{unfolding saturation,} \end{array} \right .\vspace{0.2 cm}\label{eeee}\end{equation}
assigning the (non-differential) evolution constitutive equation.

\vskip1cm

\noindent {\it Remark 4.} 
Observe that the constitutive behavior of the continuum model is assigned through  three  only parameters: the parameter $\mu_o$ in (\ref{muo}) defining the infinitesimal shear modulus, and the parameters $ \lambda_{\mbox{\tiny {\it rel}}}^{\mbox{\tiny {\it un}}}$ in (\ref{luuu}) and  $I_c^o$ in (\ref{Ioc}). These two last parameters regulate the initial elastic limit $I^o=( \lambda_{\mbox{\tiny {\it rel}}}^{\mbox{\tiny {\it un}}}\,)^2 I_c^o$ and (see {\it e.g.} Fig.\ref{SE}) the response coefficient $\aleph$. In particular we remark (see  (\ref{IcIc})) the fundamental role of $\lambda_{\mbox{\tiny {\it rel}}}^{\mbox{\tiny {\it un}}}$ (see again Fig.\ref{SE}) defining the response coefficient $\bar \aleph$ with $\displaystyle\lim_{\lambda_{\mbox{\tiny {\it rel}}}^{\mbox{\tiny {\it un}}}\rightarrow 1}\bar \aleph=+\infty$. Finally, we point out that according with (\ref{FpFp}) these parameters regulate also the residual stretches that decrease as  $\lambda_{\mbox{\tiny {\it rel}}}^{\mbox{\tiny {\it un}}}$ and $I_c^o$ grow.
\vspace{0.6 cm}

To discuss  the thermodynamical consistence of the model, we observe that the Clausius-Duhem inequality requires the positivity of the rate of energy density production per unit volume in the reference configuration ${\cal B}_o$ \cite{CG}
$$\gamma=
(\det \mbox{\boldmath{$F$}})\,\,  \mbox{\boldmath{$T$}}\cdot \mbox{\boldmath{$L$}}- (\varphi_e(\mbox{\boldmath{$F$}}_e,I_c) \det \mbox{\boldmath{$F$}}_p)\dot { } \geq 0,$$
where $\mbox{\boldmath{$L$}}=\dot{\mbox{\boldmath{$ F$}}}\mbox{\boldmath{$ F$}}^{-1}$ is the velocity gradient.
We have
$$(    \varphi_e\det \mbox{\boldmath{$F$}}_p)\dot{} =\det \mbox{\boldmath{$F$}}_p\left ( \varphi_e  \mbox{\boldmath{$F$}}_p^{-T} \cdot \dot{\mbox{\boldmath{$F$}}}_p +  \partial_{ \mbox{\boldmath{\tiny $F$}}_e}\varphi_e \cdot  \dot {\mbox{\boldmath{$F$}}}_e - G \,\dot I_c \right )$$
where we used (\ref{defdef}) and (\ref{ddff}). Here 
\begin{equation}\begin{array}{l}ÊG:= \partial_{ I_c}\varphi_e(\mbox{\boldmath{$F$}}_e,I_c)=\bar g(I_e,I_e^c)-\bar g(3,I_e^c)\geq 0\vspace{0.2 cm }\end{array}
  \end{equation}  represents the driving force measuring the dissipation associated to the variation of the internal variable $I_c$ and $\bar g(I_e,I_e^c)= \frac{\mu_o}
   {4} \frac{\sqrt{\frac{I^e}{I^e_c}}}{\left ( 1- \sqrt{\frac{I^e}{I^e_c}}\right )^2}$ .
On the other hand, due to (\ref{FF}) we have
$$\mbox{\boldmath{$L$}}=\dot {\mbox{\boldmath{$F$}}}_e\mbox{\boldmath{$F$}}_e^{-1}+\mbox{\boldmath{$F$}}_e \dot {\mbox{\boldmath{$F$}}}_p\mbox{\boldmath{$F$}}_p^{-1}\mbox{\boldmath{$F$}}_e^{-1},
$$
so that, since $\det \mbox{\boldmath{$F$}}=\det \mbox{\boldmath{$F$}}_p$, we obtain
\begin{equation}\label{CDCD}\frac{\gamma}{\det\mbox{\boldmath{$F$}}_p}=G \dot I_c+(\mbox{\boldmath{$T$}}\mbox{\boldmath{$F$}}_e^{-T}-\partial_{ \mbox{\boldmath{\tiny $F$}}_e}\varphi_e  )\cdot \dot{\mbox{\boldmath{$F$}}}_e-  \mathbb{E} \cdot \dot{\mbox{\boldmath{$F$}}}_p\mbox{\boldmath{$F$}}_p^{-1}. \end{equation}
Here 
$$ \mathbb{E}= \varphi_e \mbox{\boldmath{$I$}}-\mbox{\boldmath{$F$}}_e^T \mbox{\boldmath{$T$}}
\mbox{\boldmath{$F$}}_e^{-T}$$
is the {\it Eshelby tensor} working for the volumetric growth due to unfolding (see {\it e.g.} \cite{AG}).
Then, using (\ref{equilb}) and (\ref{kkk}) we obtain that  the second term in (\ref{CDCD}) is null, so that we deduce the {\it reduced dissipation inequality}
$$G \dot I_c -  \mathbb{E}\cdot \dot{\mbox{\boldmath{$F$}}}_p\mbox{\boldmath{$F$}}_p^{-1}  \geq0.$$
Observe that using (\ref{Fp}) we obtain
$$- \mathbb{E}\cdot \dot{\mbox{\boldmath{$F$}}}_p\mbox{\boldmath{$F$}}_p^{-1} =(\mbox{tr} \mbox{\boldmath{$T$}} - 3 \varphi_e)\frac{\dot I_c}{4 I_c}$$
and the reduced dissipation inequality can be rewritten as  
\begin{equation}\label{rrddii} \left(G
+\frac{\mbox{tr} \mbox{\boldmath{$T$}} - 3 \varphi_e}{4 I_c} \right )\dot I_c\geq0.\vspace{0.2 cm}\end{equation}
In the numerical calculations of the following section we verify that the condition (\ref{rrddii}) is satisfied, thus granting the thermodynamical consistence of the obtained responses.

In closing, we may observe that for $\dot I_c\geq 0$
$$\mbox{tr} \mbox{\boldmath{$T$}} \geq 3 \varphi_e$$
represents a sufficient condition to respect the Clausius-Duhem inequality. This condition requires that the rate of elementary work due to the volume growth is higher than the corresponding rate of energy increase.

\vspace{0.3 cm}

\noindent {\it Remark 5}.\hspace{0.2 cm} The dissipation potential for the three-dimensional continuum model is assigned, based on (\ref{rrddii}), by the expression
$$D=D(I,I_c,\dot I_c)= \left(G
+\frac{\mbox{tr} \mbox{\boldmath{$T$}} - 3 \varphi_e}{4 I_c} \right )_{
{ | }_{I=I^{max}}}\hspace{- 1.2 cm}\dot I_c.\vspace{0.2 cm}$$
 \vspace{0.15 cm}

\noindent {\it Remark 6}.\hspace{0.2 cm} With reference to the existence problem of the elastic solution at fixed $I_c$, since $\partial \varphi^e(I^e,\nu)/\partial I^e>0$ and  $\partial^2 \varphi^e(I^e,\nu)/\partial^2 I^e>0$, we obtain that the elastic energy is a polyconvex function  (see \cite{Sh}) of the elastic strain tensor $\mbox{\boldmath{$F$}}^e$. 
\vspace{0.5 cm}

\section{Numerical examples and experimental comparison}\label{DG}

In this section, in order to show the feasibility of the obtained constitutive model in describing the main experimental effects of biological materials, we analyze the deformation classes of uniaxial extension, biaxial extension and shear. The quantitative comparison with cyclic experimental testing of some biological materials is also shown.

\subsection{Uniaxial extension}
First, consider the simple case of uniaxial extension  in the direction $\mbox{\boldmath{$e$}}_1$ with $\mbox{\boldmath{$F$}}^e=
\lambda^e\, \mbox{\boldmath{$e$}}_1 \otimes \mbox{\boldmath{$e$}}_1+ \frac{1}{\sqrt{\lambda^e} }\,( \mbox{\boldmath{$e$}}_2 \otimes \mbox{\boldmath{$e$}}_2 + 
 \mbox{\boldmath{$e$}}_3 \otimes \mbox{\boldmath{$e$}}_3) $, satisfying the kinematic constraint (\ref{kkk}).
 Thus by (\ref{FpFp}) and (\ref{FF}) we obtain
$$\mbox{\boldmath{$B$}}=\left [ \begin{array}{ccc}
\lambda^2& 0 & 0 \\
0 &\left ( \frac{I_c}{I_c^o} \right )^{\frac{3}{4}}  \lambda^{-1} & 0\\
0 &0&\left ( \frac{I_c}{I_c^o} \right )^{\frac{3}{4}}  \lambda^{-1} \end{array} \right ],
$$ where  $\lambda=(\frac{I_c}{I_c^o})^{\frac{1}{4} }\lambda^e$ represents the stretch in the $\mbox{\boldmath{$e$}}_1$ direction. As a result we obtain 
 
$$I=\lambda^2+2\left ( \frac{I_c}{I_c^o} \right )^{\frac{3}{4}}  \frac{1}{\lambda}.$$

\begin{figure}[!h]
\begin{center}\hspace{- 1 cm}$$
\begin{array}{cc}\includegraphics[scale=0.6]{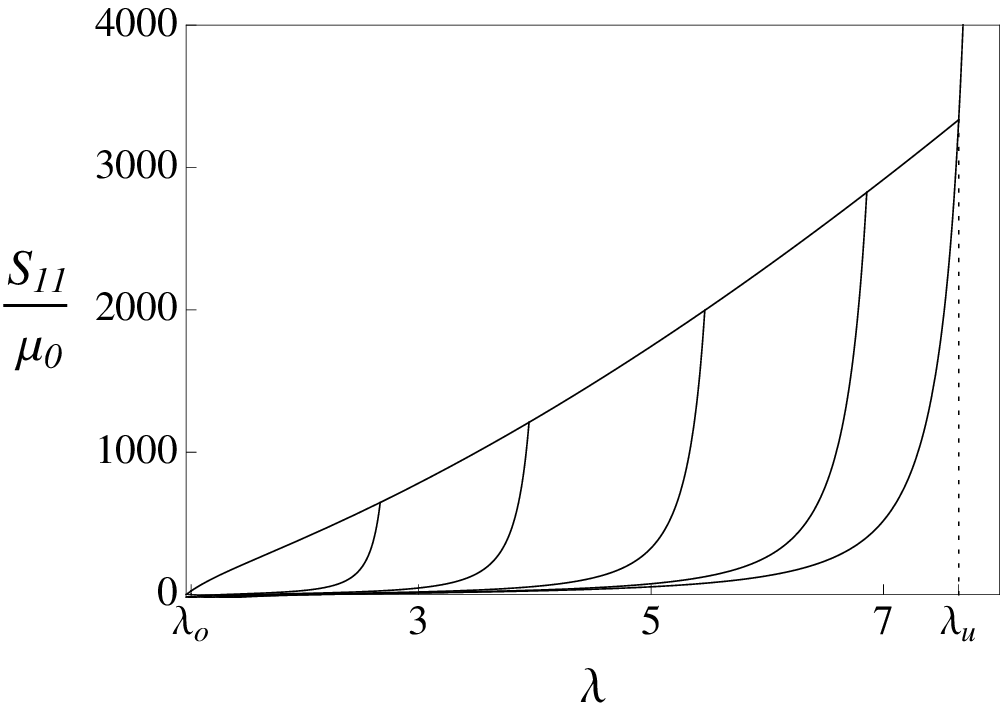}a) &\hspace{- 0.3 cm} \includegraphics[scale=0.6]{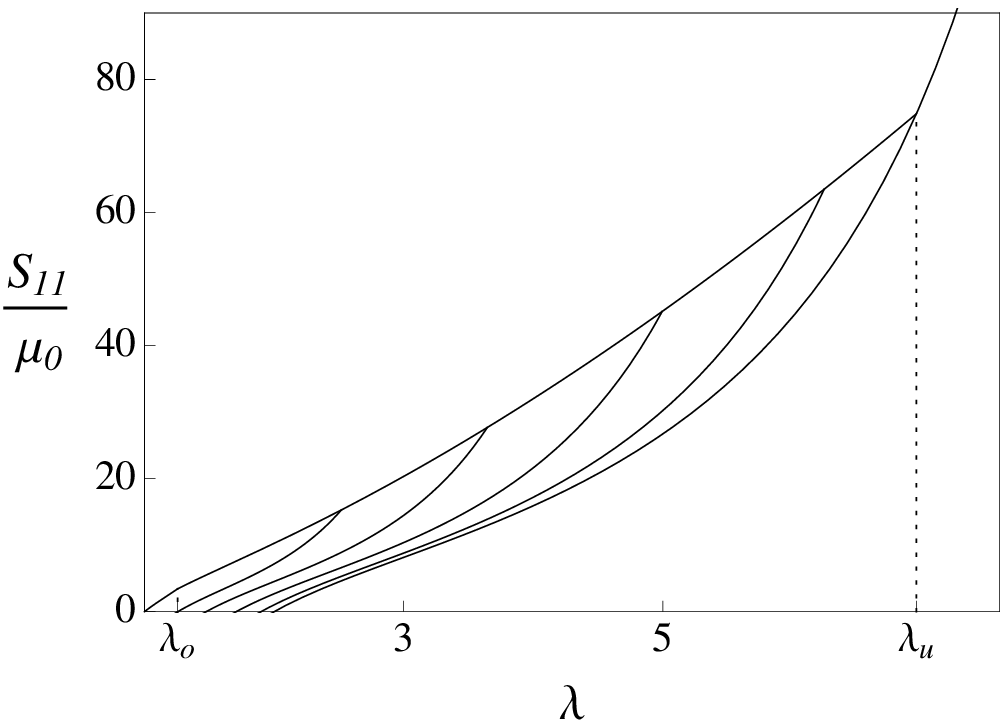}b)\\
\includegraphics[scale=0.6]{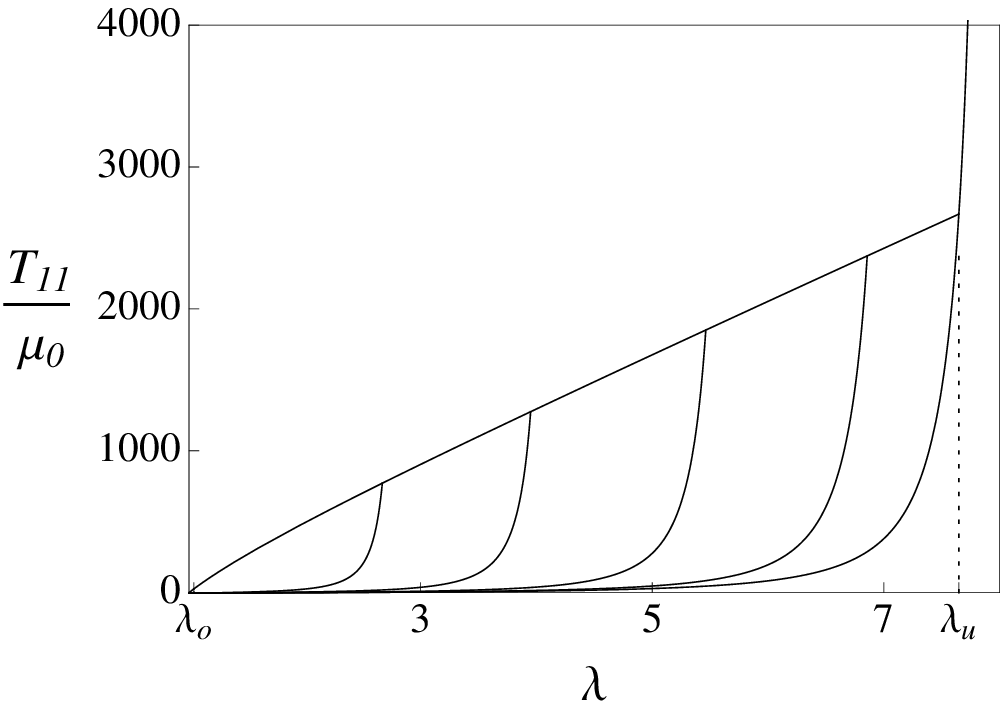} c) &\hspace{- 0.3 cm} \includegraphics[scale=0.6]{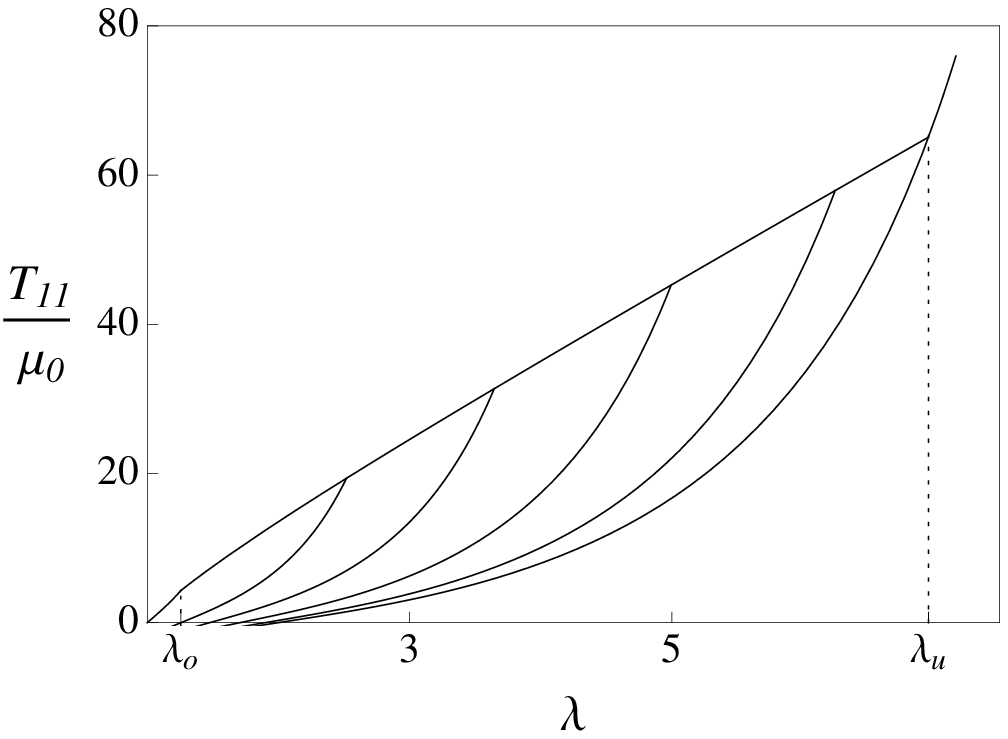}d) \end{array}$$
\caption{\label{SE} Simple extension. In a) and b) we represent the Piola stress and in c) and d)  the Cauchy stress as a function of the stretch. In a) and c) we assume  $I_c^0=3.3$, $I_c^1=55$, $\lambda_{\mbox{\tiny {\it rel}}}^{\mbox{\tiny {\it un}}} = 0.95$ in b) and d) $I_c^0=7.5$, $I_c^1=120$, $\lambda_{\mbox{\tiny {\it rel}}}^{\mbox{\tiny {\it un}}} = 0.65$. Here $\lambda_o$ and $\lambda_u$ are the stretches corresponding to the initiation of unravelling ($I=I^o$) and to unravelling saturation ($I=I^1$), respectively.
}
\end{center}
\end{figure}

By imposing $\mbox{\boldmath{$T$}}_{22}=\mbox{\boldmath{$T$}}_{33}=0$, we determine the pressure 
$$p=\aleph \left (\frac{I_c}{I_c^o} \right )^{\frac{3}{4}} \, \frac{1}{\lambda}, $$

\noindent so that the stress-strain relation is given by
$$\mbox{\boldmath{$T$}}_{11}(\lambda)=\aleph \left (  \lambda ^2- \left (\frac{I_c}{I_c^o} \right )^{\frac{3}{4}}\, \frac{1}{\lambda} \right )$$
or in terms of the Piola stress by
$$\mbox{\boldmath{$S$}}_{11}(\lambda)=\aleph \left (  \lambda - \left (\frac{I_c}{I_c^o} \right )^{\frac{3}{4}}\, \frac{1}{\lambda^2} \right ) \left (\frac{I_c}{I_c^o} \right )^{\frac{3}{4}}.$$
In Fig.\ref{SE} we represent the Piola (nominal) and Cauchy (true) stresses as a function of the cyclically varying total stretch $\lambda$, deduced by the evolution equations (\ref{eeee}). In order to show the fundamental role of the parameter $\lambda_{\mbox{\tiny {\it rel}}}^{\mbox{\tiny {\it un}}}$,  in the figure we consider two different values of this parameter. By decreasing  $\lambda_{\mbox{\tiny {\it rel}}}$ we may observe an increase of residual stretches and a decrease of the slope of the equilibrium curves.

\subsection{Biaxial extension}

Consider now the case of biaxial extension in the plane $\mbox{\boldmath{$e$}}_1$-$\mbox{\boldmath{$e$}}_2$ with elastic deformation tensor
$\mbox{\boldmath{$F$}}^e=
\lambda^e \, (\mbox{\boldmath{$e$}}_1 \otimes \mbox{\boldmath{$e$}}_1+ \mbox{\boldmath{$e$}}_2 \otimes \mbox{\boldmath{$e$}}_2) +\frac{1}{{\lambda^e}^2} \,
 \mbox{\boldmath{$e$}}_3 \otimes \mbox{\boldmath{$e$}}_3 $, respecting the kinematic constraint (\ref{kkk}).By (\ref{FpFp}) and (\ref{FF}) we obtain

$$\mbox{\boldmath{$B$}}=\left [ \begin{array}{ccc}
\lambda^2& 0 & 0 \\
0 &\lambda^2& 0\\
0 &0&\left ( \frac{I}{I_c^o} \right )^{\frac{3}{2}} \lambda^{-4} \end{array} \right ],
$$
with $\lambda=(\frac{I_c}{I_c^o})^{\frac{1}{4} }\lambda^e$ representing the in-plane stretch; thus $$I=2 \lambda^2+\left ( \frac{I}{I_c^o} \right )^{\frac{3}{2}}\frac{1}{\lambda^4}.$$

The condition $\mbox{\boldmath{$T$}}_{33}=0$ gives the pressure 
$$p=\aleph\left ( \frac{I}{I_c^o} \right )^{\frac{3}{2}}\frac{1}{\lambda^4}, $$

\begin{figure}[!h]
\begin{center}$$
\begin{array}{rr}\includegraphics[scale=0.65]{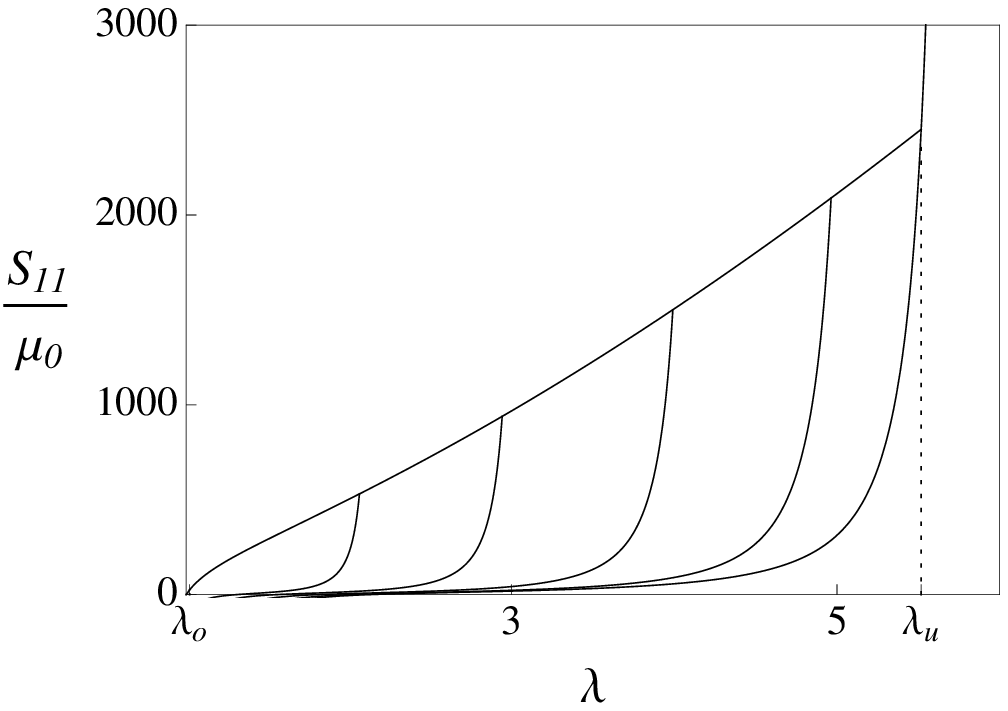} &\includegraphics[scale=0.65]{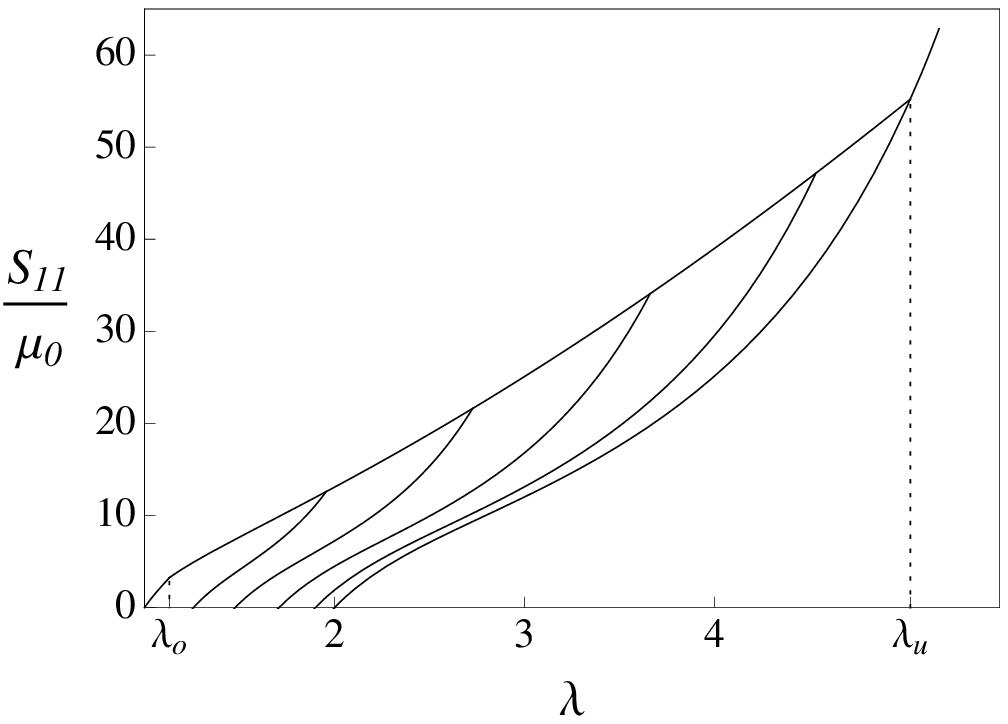} \\ a) & b)
\end{array}$$
\caption{\label{BE} Biaxial extension: Piola stress as a function of the stretch. In a) $I_c^0=3.3$, $I_c^1=55$, $\lambda_{\mbox{\tiny {\it rel}}}^{\mbox{\tiny {\it un}}} = 0.95$, in b) $I_c^0=7.5$, $I_c^1=120$, $\lambda_{\mbox{\tiny {\it rel}}}^{\mbox{\tiny {\it un}}} = 0.65$. Here $\lambda_o$ and $\lambda_u$ are the stretches corresponding to the initiation of unravelling ($I=I^o$) and to unravelling saturation ($I=I^1$), respectively.
}
\end{center}
\end{figure}

\noindent so that  the stress-strain relation are  given by
$$\mbox{\boldmath{$T$}}_{11}(\lambda)=\aleph \left ( \lambda ^2-\left ( \frac{I}{I_c^o} \right )^{\frac{3}{2}}\, \frac{1}{\lambda^4} \right )$$
 and
 $$\mbox{\boldmath{$S$}}_{11}(\lambda)= \left (\frac{I_c}{I_c^o} \right )^{\frac{3}{4}}\aleph \left ( \lambda -\left ( \frac{I}{I_c^o} \right )^{\frac{3}{2}}\, \frac{1}{\lambda^5} \right ).$$
In Fig.\ref{BE} we represent the Piola stress for two different values of  $\lambda_{\mbox{\tiny {\it rel}}}^{\mbox{\tiny {\it un}}}$.

\subsection{Shear strain}

\begin{figure}[!h]
\begin{center}$$
\begin{array}{rr}\includegraphics[scale=0.65]{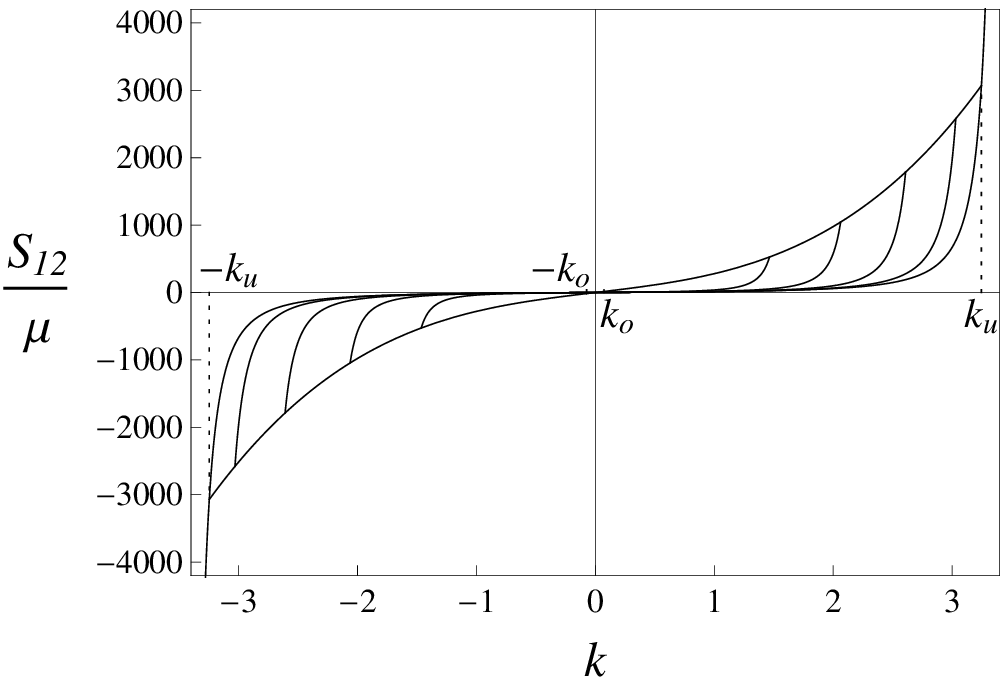} &\includegraphics[scale=0.65]{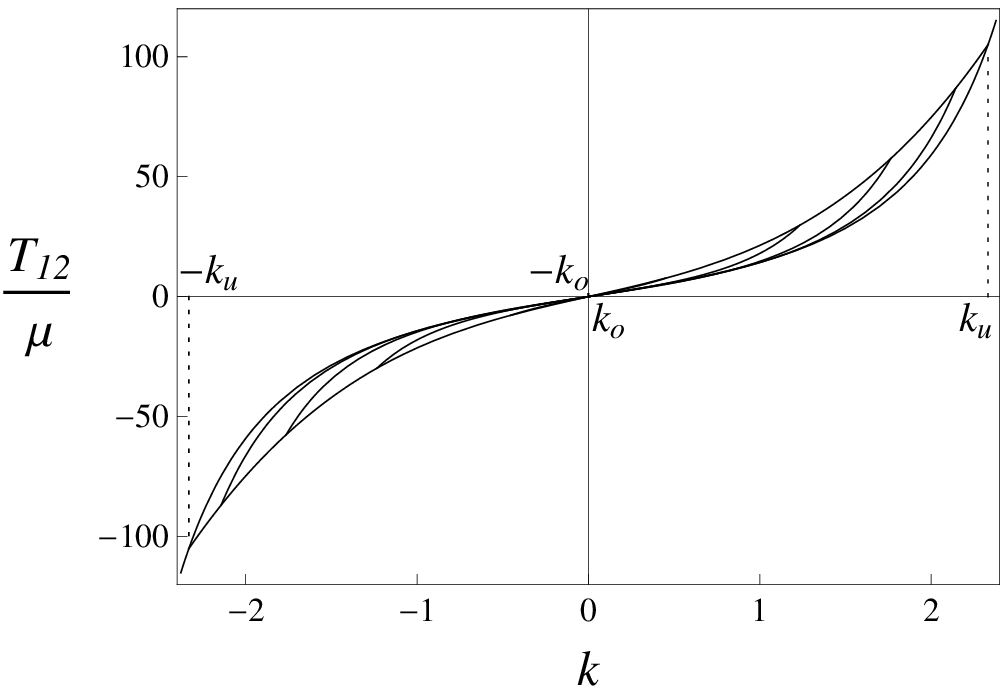} \\
a) & b)\end{array}$$
\caption{\label{SS} Shear deformations: Piola stress as a function of the shear stretch.  Here $I_c^0=3.3$, $I_c^1=68$, in a) $\lambda_{\mbox{\tiny {\it rel}}}^{\mbox{\tiny {\it un}}} = 0.95$, in b) $\lambda_{\mbox{\tiny {\it rel}}}^{\mbox{\tiny {\it un}}} = 0.75$. In the figure $k_o$ and $k_u$ are the shears corresponding to the initiation of unravelling ($I=I^o$) and to unravelling saturation ($I=I^1$), respectively.
}
\end{center}
\end{figure}

Consider finally the case of shear strain $\mbox{\boldmath{$F$}}^e=\mbox{\boldmath{$I$}}+k \,
 \mbox{\boldmath{$e$}}_1 \otimes \mbox{\boldmath{$e$}}_2 $, so that by  (\ref{FpFp}) and (\ref{FF}) we obtain
$$\mbox{\boldmath{$B$}}=\left ( \frac{I_c}{I_c^o} \right )^{\frac{3}{2}}\left [ \begin{array}{ccc}
1 + k^2 & k & 0 \\
k & 1& 0\\
0 &0&1 \end{array} \right ],
$$
with $k$ representing the  total shear strain (coinciding with the elastic one).
Thus we have
$$I=\left ( \frac{I_c}{I_c^o} \right )^{\frac{3}{2}}(3+ k^2).$$
As a result  the stress-strain relation is  given by

$$\mbox{\boldmath{$T$}}_{12}(k)=\aleph \left ( \frac{I_c}{I_c^o} \right )^{\frac{1
}{2}}\, k$$
or in terms of Piola stress by
$$\mbox{\boldmath{$S$}}_{12}(k)=\aleph \left ( \frac{I_c}{I_c^o} \right )\, k.$$
In Fig.\ref{SS} we represent the Piola stress for two different values of  $\lambda_{\mbox{\tiny {\it rel}}}^{\mbox{\tiny {\it un}}}$.
Observe that, under our hypothesis of isotropic residual stretch ({\it i.e.} spherical growth tensor $\mbox{\boldmath{$F$}}^p$),  the shear stress-strain relation does not exhibit residual strain. This result is also supported by the experimental observations in the following section.

\subsection{Experimental validation}

\begin{figure}[!h]
$$\begin{array}{rr} \hspace{-1.13 cm}   \includegraphics[scale=0.3]{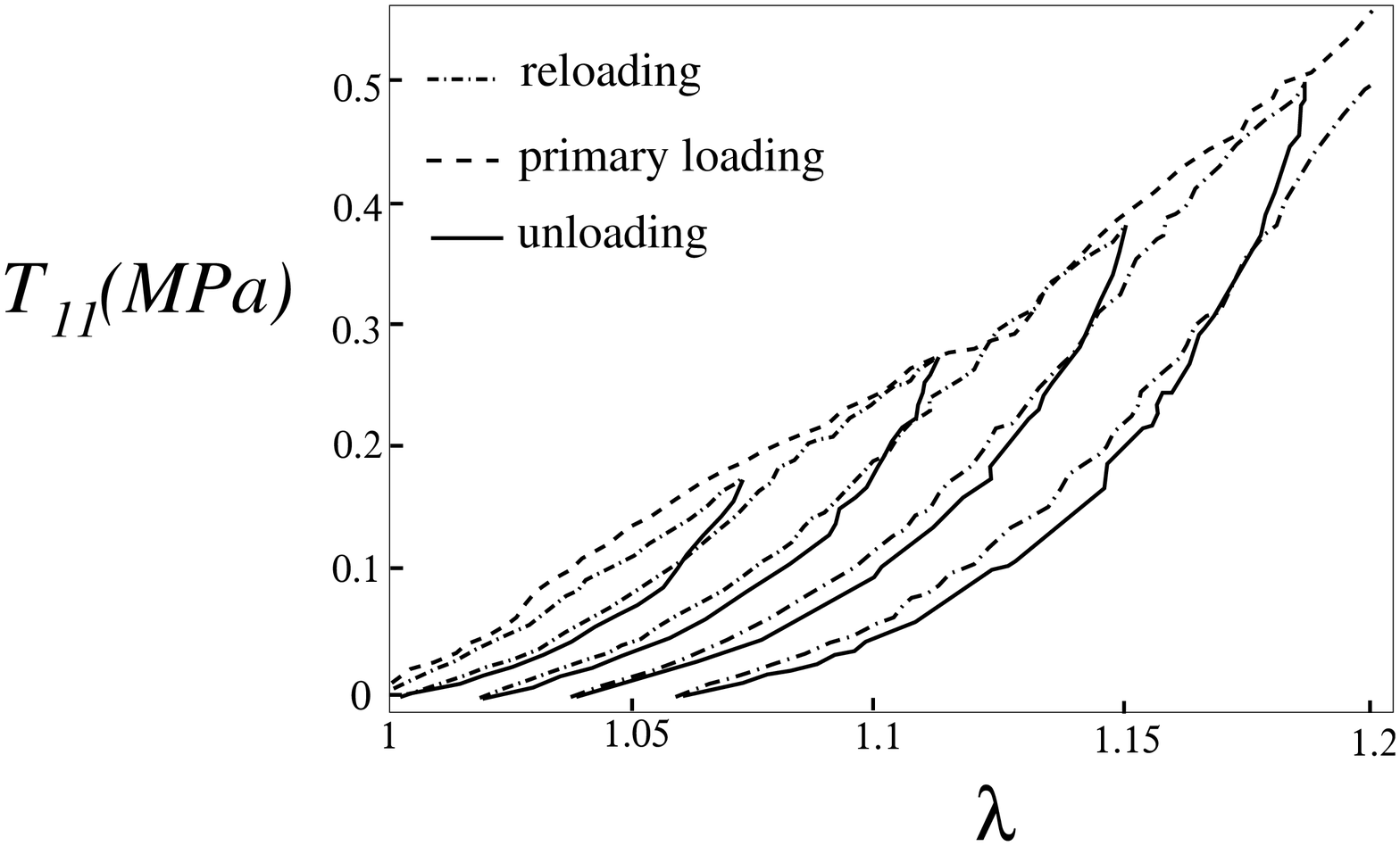}  & \hspace{-0.53 cm}  \includegraphics[scale=0.7]{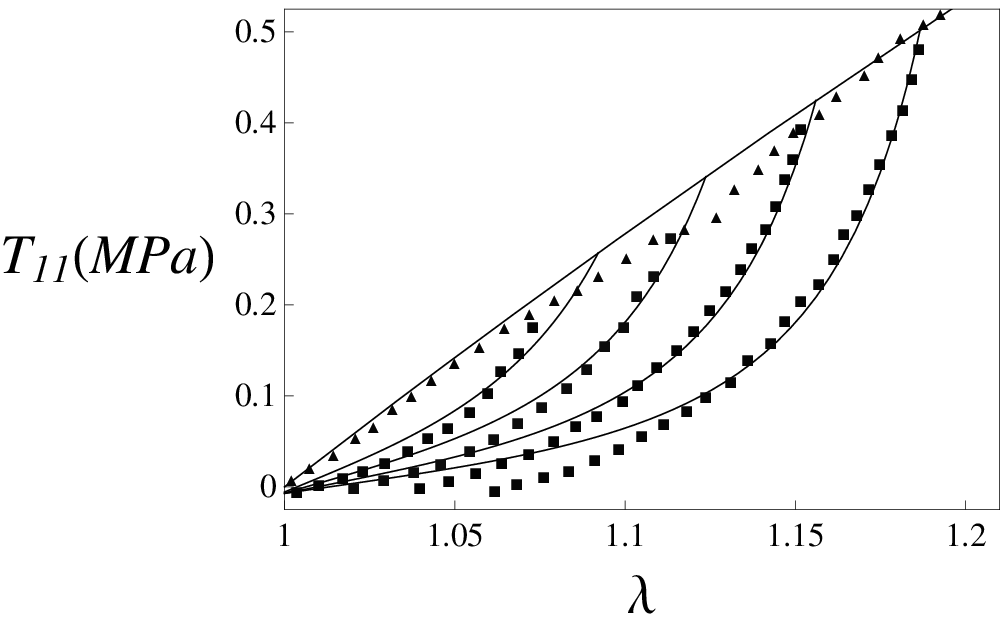} \\ a) & b) \vspace{0.5 cm} \end{array}$$
\begin{center}\caption{\label{MS} In a) we reproduce the experiments 
of \cite{Dobl} of cyclic loading of mouse skin. 
In b) we reproduce the experimental results with our theoretical model: here  $\mu_o= 0.18$ kPa, $I_c^0=3.33$,  $\lambda_{\mbox{\tiny {\it rel}}}^{\mbox{\tiny {\it un}}} = 0.99032$.}
\end{center}
\end{figure}

The first biological material we consider is the {\it skin of mice} whose experimental behavior reported in \cite{Dobl} is reproduced in  Fig.\ref{MS}$_a$. A theoretical analysis in the pseudoelaticity  framework of the hysteretic behavior of this biological tissue   has been recently proposed in \cite{Zun}.  In the figure we represent by square dots the primary loading curve, by triangular dots the unloading curves, and by circular dots the reloading ones.  As obtained in the macroscopic response of the proposed theoretical scheme,  we observe the softening behavior ({\it i.e.} the stress decreases with the maximum attained elongation), the residual stretches under unloading, the Return Point Memory property ({\it i.e.} the system reconnects to the unloading point on the primary loading curve under reloading). Also we may observe that, coherently with the theoretical results, the stiffness of the reloading curves reduces as the maximum  assigned stretch grows, showing an increasing concavity and increasing maximum tangent modulus approaching the primary loading curve. 

\begin{figure}[!h]\begin{center}
\includegraphics[scale=0.7]{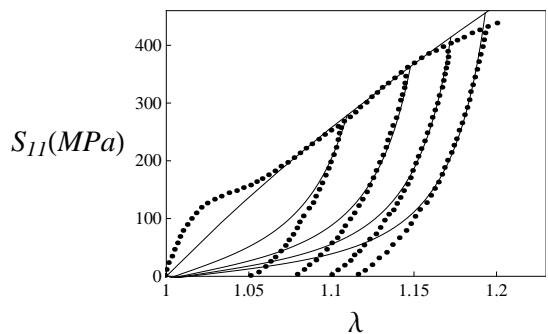} \caption {Dotted lines: experimental behavior of Senegalensis Nephila spider silks in \cite{VGS}. Continuous lines: theoretical prediction for a model with  $\mu_o=5$ kPa, $I_c^0=3.042$, $\lambda_{\mbox{\tiny {\it rel}}}^{\mbox{\tiny {\it un}}} = 0.993$. }\label{expexpexp}
\end{center}
\end{figure}

In Fig.\ref{expexpexp}$_b$
we show the capability of our model to quantitatively reproduce the experimental curves and their main effects described above. We remark that, as in the other experiment reported in Fig.\ref{expexpexp}$_c$, we may observe that the theoretical model underestimates the presence of residual stretches. We believe that this result is due to our assumption of uniform residual stretch of the chains: roughly speaking we underestimate the residual stretch because we identifies its value in the loading direction with the mean value.

In Fig.\ref{expexpexp}$_c$ we predict by our model the experimental behavior of {\it Nephila senegalensis spider silks} reported in \cite{VGS}.
The softening and dissipative behavior of this material has been analyzed 
in \cite{DeTommasi-Puglisi-Saccomandi2}  based on a phenomenological damage model taking care of the stretch-induced variation of the contour length. Observe again that this material shows the main physical effects described above and that they are captured by our model. 
Once more we remark that the model underestimates the residual stretch and that by considering anisotropic effects both residual stretches and the behavior at low stress could be better described.

\begin{figure}[!h]
\begin{center} \includegraphics[scale=0.7]{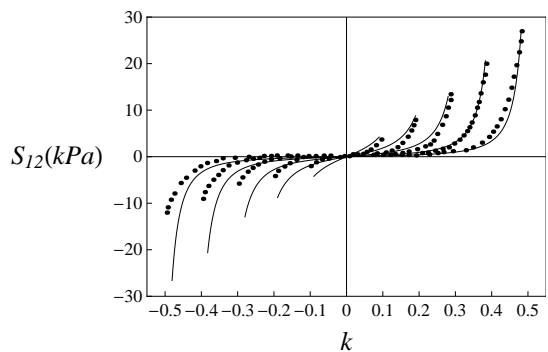}\caption {Dotted lines: experimental behavior of pig (passive) miocardium tissues
reported in \cite{DSY}. Continuous lines: theoretical prediction for a system with  $\mu_o= 0.095$ MPa, $I_c^0=3.06$, $\lambda_{\mbox{\tiny {\it rel}}}^{\mbox{\tiny {\it un}}} = 0.993$. }\label{expexpexpb}
\end{center}
\end{figure}

In Fig.\ref{expexpexpb} we represent the shear test for {\it pig (passive) myocardium tissues}
reported in \cite{DSY}. The figure shows the possibility of well describing the experimental behavior also in the case of shear deformations. Observe that, as remarked above, in accordance with the theoretical prediction, in the case of shear the experiments show negligible residual deformations. The experiments exhibit an `unusual' asymmetric behavior that does not permit a well prediction of the negative part of the curves. 

\newpage
\section{Conclusions} 

We developed a multiscale  approach to deduce the three-dimensional response of  tissues and networks of macromolecules with unfolding domains starting from the mechanical behavior of the single chain. 
Our approach is fully analytical and the obtained constitutive model depends explicitly on few material parameters with a clear physical interpretation and an analytical derivation from microscopic properties. As a result of our  approach, we obtain an interpretation of the experimentally observed residual stretches and stresses, here deduced as an effect of the variable natural configurations of the macromolecules. Our three-dimensional continuum model may be inscribed within the modern framework of Growth Mechanics with the energy density  defined in a variable natural configuration, depending on the percentage of unfolded domains. In particular, the energy is a function of the elastic strain tensor according with a Gent-type law, with a variable limit threshold of the first strain invariant. 

The comparison with the experimental behavior shows the ability of the model to describe  the main phenomena observed in the cyclic response of biological materials with unfolding macromolecules. We believe that the possibility of analytically relating the macroscopic material response of these materials to the microstructure parameters represents an important step in the  comprehension, modeling and design of macromolecular materials. 

Our approach is  general and several augmentation of the model can be considered. At the scale of the single chain different models (such as Freely Jointed Chains model for polymeric materials)
for the entropic and for the enthalpic energy ({\it e.g.} inhomogeneous properties of the unfolding domains with variable energy barrier and contour length) can be introduced. At the macroscale, where we showed the efficiency of the Arruda-Boyce scheme in describing isotropic behaviors, other models can be considered such as the three-chain cell structure (allowing the modeling of anisotropic effects) or possible extensions taking care of non-affine effects \cite{K}. The possibility of modeling anisotropic effects is in our opinion fundamental for  a more detailed description of the residual stretch effect (underestimated by the homogeneity assumptions of the Arruda-Boyce approach). On the other hand the introduction of variable unfolding energy barriers can be  crucial to describe the important hardening effect observed for some materials on the primary loading curve. Finally we observe that  the cyclic experiments on biological materials can show (even important) internal hysteresis cycles that are neglected in this paper. In this perspective, by mimicking the approach proposed in \cite{DD}, we may  reproduce the internal hysteresis effect by considering a healing phenomenon induced by a partial refolding of the crystals upon unloading. These extensions will be the object of our future research.

\vspace{1 cm}

\noindent {\bf Acknowledgments}.  The work of D.D. and G.P. has been supported by Progetto di ricerca industriale-Regione Puglia, ÔModelli innovativi per sistemi meccatroniciÕ and Fondi di Ricerca di Ateneo of Politecnico di Bari. G.S. is grateful to the Istituto Nazionale di
Alta Matematica (Italy) and PRIN 2009 Matematica e meccanica dei sistemi
biologici e dei tessuti molli for financial support.

\begin{section}*{Appendix: List of symbols}

$$ \begin{footnotesize}
\begin{array} {ll}
L & \mbox{current chain length} \\
L^{\mbox{\tiny {\it max}}} & \mbox{maximum attained value of } L\\
f & \mbox{chain equilibrium force} \\
f^{\mbox{\tiny {\it un}}} & \mbox{chain unfolding force} \\
L_c & \mbox{chain contour length} \\
L_c^{\mbox{\tiny 1}} & \mbox{totally unfolded chain contour length} \\
L_c^{\mbox{\tiny 0}} & \mbox{initial chain contour length}\\
L^{\mbox{\tiny {\it un}}} & \mbox{chain unfolding length}\\
l_c & \mbox{single domain contour length}\\
n & \mbox{current number of folded domains}\\
n_t & \mbox{total number of folded domains}\\
n_o & \mbox{initial number of folded domains}\\
P& \mbox{chain persistence length}\\
k_B & \mbox{Boltzmann constant} \\
Q & \mbox{single domain unfolding energy } \\
T & \mbox{temperature} \\
\lambda_{\mbox{\tiny {\it rel}}} & \mbox{chain relative stretch}\\
\lambda_{\mbox{\tiny {\it rel}}}^{\mbox{\tiny {\it un}}} & \mbox{unfolding chain relative stretch}\\
\Phi_e & \mbox{chain entropic energy}\\
\Phi & \mbox{total chain energy}\\
\nu & \mbox{fraction of unfolded domains}\\
\nu_o & \mbox{initial fraction of unfolded domains}\\
\Gamma & \mbox{chain energy dissipation rate}\\
D & \mbox{dissipation potential}\\
g & \mbox{chain driving force}\\
L_o & \mbox{initial chain natural length}\\
L_p & \mbox{present chain natural length}\\
\lambda & \mbox{chain and continuum total stretch}\\
\lambda_e & \mbox{chain and continuum elastic stretch}\\
\lambda_p & \mbox{chain and continuum permanent stretch}\\
\lambda_{\mbox{\tiny o}} & \mbox{initial unfolding stretch threshold} \\
\lambda_{\mbox{\tiny 1}} & \mbox{unfolding saturation stretch threshold } \\
\end{array} \end{footnotesize}
$$

$$ \begin{footnotesize}
\begin{array} {ll}
\lambda_c & \mbox{contour length stretch}\\
\mbox{\boldmath{$f$}}  & \mbox{deformation function}\\
{\cal B}_o  & \mbox{reference configuration}\\
{\cal B}  & \mbox{present body configuration}\\
{\cal B}_p  & \mbox{relaxed  configuration}\\
\mbox{\boldmath{$F$}} & \mbox{ deformation tensor}\\
\mbox{\boldmath{$F$}}_e & \mbox{elastic deformation tensor}\\
\mbox{\boldmath{$F$}}_p & \mbox{growth tensor}\\
\mbox{\boldmath{$B$}} & \mbox{left Cauchy-Green  tensor}\\
\mbox{\boldmath{$B$}}_e & \mbox{elastic left Cauchy-Green  tensor}\\
\lambda_i & \mbox{principal stretches}\\
\mbox{\boldmath{$e$}}_i & \mbox{principal unit vectors}\\
I & \mbox{first  invariant of } \mbox{\boldmath{$B$}}\\
I^e & \mbox{first  invariant of } \mbox{\boldmath{$B$}}_e\\
I_c & \mbox{contour length value of } I \\
I_c^e & \mbox{contour length value of } I^e \\
I_c^o & \mbox{initial value of } I_c \\
I_c^1 & \mbox{unfolding saturation threshold of } I_c\\
I^o & \mbox{initial unfolding threshold of } I  \\
I^1 & \mbox{unfolding saturation threshold of } I\\
N & \mbox{chains per unit volume in } {\cal B}_p\\
N_o & \mbox{chains per unit volume in } {\cal B}_o\\
\varphi_e & \mbox{three dimensional energy density function}\\
\mu_o & \mbox{infinitesimal shear modulus}\\
\mbox{\boldmath{$T$}}  & \mbox{Cauchy stress tensor}\\
\mbox{\boldmath{$S$}}  & \mbox{Piola stress tensor}\\
p & \mbox{pressure}\\
\gamma  & \mbox{rate of energy density production in } {\cal B}_o \\
G  & \mbox{macroscopic driving force} \\
\mbox{\boldmath{$L$}}  & \mbox{velocity gradient}\\
\mathbb{E}& \mbox{Eshelby tensor }   \\
\aleph& \mbox{response coefficient}\\
\end{array} \end{footnotesize}
$$

\end{section}

\end{document}